\documentclass[a4paper,fleqn,usenatbib]{mnras}

\usepackage{newtxtext,newtxmath}

\usepackage{subfigure}
\usepackage{epsfig}
\usepackage{rotating}
\usepackage{newtxtext,newtxmath}
\usepackage[T1]{fontenc}
\usepackage{ae,aecompl}
\usepackage{tikz}

\usepackage{graphicx}	
\usepackage{amsmath}	
\usepackage{amssymb}	

\newcommand{\Ms}{M$_{\odot}$}
\newcommand{\Ls}{L$_{\odot}$}

\title[Supermassive Primordial Stars II]{On the detection of supermassive primordial stars. II. Blue supergiants.}

\author[M. Surace et al.]{Marco Surace,$^{1}$
Erik Zackrisson,$^{2}$ 
Daniel J. Whalen,$^{1,3}$\thanks{E-mail: daniel.whalen@port.ac.uk}
Tilman Hartwig,$^{4,5,6}$
\newauthor
S.~C.~O. Glover,$^{7}$
Tyrone E. Woods,$^{8}$
Alexander Heger,$^{9,10}$
S.~C.~O. Glover$^{11}$
\\
\\
$^{1}$Institute of Cosmology and Gravitation, University of Portsmouth, Portsmouth PO1 
3FX, UK \\ 
$^{2}$Observational Astrophysics, Department of Physics and Astronomy, Uppsala University, Box 516, SE-751 20 Uppsala, Sweden \\
$^{3}$Ida Pfeifer Professor, University of Vienna, Department of Astrophysics, Tuerkenschanzstrasse 17, 1180, Vienna, Austria \\
$^{4}$Kavli IPMU (WPI), UTIAS, The University of Tokyo, Kashiwa, Chiba 277-8583, Japan \\
$^{5}$Department of Physics, School of Science, University of Tokyo, Bunkyo, Tokyo 113-0033, Japan \\
$^{6}$Institute for Physics of Intelligence, School of Science, The University of Tokyo, Bunkyo, Tokyo 113-0033, Japan \\
$^{7}$Universit\"at Heidelberg, Institut f\"ur Theoretische Astrophysik, Albert-Ueberle-Str. 2, 69120 Heidelberg, Germany \\
$^{8}$Institute of Gravitational Wave Astronomy and School of Physics and Astronomy, University of Birmingham, Edgbaston, \\
Birmingham B15 2TT, United Kingdom \\
$^{9}$Monash Centre for Astrophysics, School of Physics and Astronomy, Monash University, VIC 3800, Australia \\
$^{10}$Tsung-Dao Lee Institute, Shanghai 200240, China
$^{11}$Universit\"at Heidelberg, Institut f\"ur Theoretische Astrophysik, Albert-Ueberle-Str. 2, 69120 Heidelberg, Germany
}

\date{Accepted XXX. Received YYY; in original form ZZZ}

\pubyear{2019}

\begin{document}
\label{firstpage}
\pagerange{\pageref{firstpage}--\pageref{lastpage}}
\maketitle

\begin{abstract}

Supermassive primordial stars in hot, atomically-cooling haloes at $z \sim$ 15 - 20 may have given birth to the first quasars in the universe. Most simulations of these rapidly accreting stars suggest that they are red, cool hypergiants, but more recent models indicate that some may have been bluer and hotter, with surface temperatures of 20,000 - 40,000 K. These stars have spectral features that are quite distinct from those of cooler stars and may have different detection limits in the near infrared (NIR) today. Here, we present spectra and AB magnitudes for hot, blue supermassive primordial stars calculated with the TLUSTY and CLOUDY codes.  We find that photometric detections of these stars by the {\em James Webb Space Telescope} ({\em JWST}) will be limited to $z \lesssim$ 10 - 12, lower redshifts than those at which red stars can be found, because of quenching by their accretion envelopes. With moderate gravitational lensing, {\em Euclid} and the {\em Wide-Field Infrared Space Telescope} ({\em WFIRST}) could detect blue supermassive stars out to similar redshifts in wide-field surveys.  

\end{abstract}


\section{Introduction}

Supermassive stars (SMSs) have long been the subject of analytical studies \citep[e.g.,][]{iben63,chandra64,fowler64,fowler66} and numerical simulations \citep[e.g.,][]{af72a,st79,fuller86,baum99,sun18,but18}. But credible scenarios for their formation have only recently been found:  supermassive primordial star (SMS) formation in atomically-cooling primordial haloes at high redshifts exposed to either unusually strong Lyman-Werner (LW) UV fluxes \citep{latif14,agarw15,chon17b,wise19} or highly supersonic baryon streaming motions \citep{lns14,hir17,srg17} or the formation of stars powered by self-annihilation of dark matter \citep['dark stars';][]{spoly08,freese08b}. Strong UV backgrounds or streaming motions can suppress star formation in a halo until it reaches masses of $\sim$ 10$^7$ \Ms\ and virial temperatures of $\sim$ 10$^4$ K that trigger rapid atomic cooling that leads to catastrophic baryon collapse that can build up a star at rates of up to $\sim$ 1 \Ms\ yr$^{-1}$ \citep{ln06,wta08,rh09b,iot14,latif15b}. Such stars may have been the origin of the first quasars, a few of which have now been discovered at $z >$ 7 \citep{mort11,ban18,smidt18}.  

Stellar evolution models indicate that primordial (Pop III) stars growing at these rates can reach masses of a few 10$^5$ \Ms\ before, in most cases, collapsing to black holes \citep[direct collapse black holes, or DCBHs;][]{um16,tyr17,hle18,hle19}. A few non-accreting Pop III SMSs may explode as thermonuclear transients \citep{montero12,wet12a,jet13a,wet13b,wet13d,chen14b}. Pop III SMSs are the leading candidates for the origin of the earliest supermassive black holes (SMBHs) because the environments of ordinary Pop III star BHs are hostile to rapid growth \citep{wan04,awa09,wf12,srd18}. In contrast, DCBHs are born with much larger masses and in much higher densities in host galaxies capable of retaining their fuel supply even when it is heated by X-rays \citep{jet13}. But Pop III star BHs, in principle, could reach large masses by super- or hyper-Eddington growth if there is enough gas to fuel their rapid growth (\citealt{mhd14,vsd15,pez16,inay16} -- see \citealt{may15,may19} for other pathways to the formation of these quasars and \citealt{rosa17,titans} for recent reviews on the first quasars).

Most studies have found that rapidly accreting Pop III stars evolve as cool, red hypergiants along the Hayashi limit, with surface temperatures of 5,000 - 10,000 K due to H$^-$ opacity in their atmospheres, at least until they reach $\sim$ 10$^5$ \Ms\ \citep{hos13}. \citet{hle17} found that SMSs can remain cool even above these masses and reach luminosities $\gtrsim 10^{10}$ \Ls. But \citet{tyr17} found that SMSs evolving from similar initial conditions quickly settle onto  hotter, bluer tracks with temperatures of 20,000 - 40,000 K. \citet{hle17} found that Pop III SMSs accreting at low rates ($\lesssim$ 0.005 \Ms\ yr;$^{-1}$) also evolve along blue tracks, as may stars with clumpy accretion due to fragmentation or turbulence in the accretion disk (\citealt{sak15} -- but see \citealt{sak16b}). Whether these differences are due to opacities, code physics (such as the numerical treatment of convection), accretion physics and boundary conditions, or numerical resolution remains unknown.

What are the prospects for observing blue Pop III SMSs today? \citet{jlj12a} semi-analytically examined the spectral features of similar stars and predicted that they would be characterized by strong Balmer emission and the conspicuous absence of Ly$\alpha$ lines due to absorption by their envelopes. The source of this flux was the hypercompact H II region of the star, whose ionising UV was trapped close to its surface by the density and ram pressure of the inflow \citep[which has also found to be true in cosmological simulations of highly-resolved atomically-cooled haloes;][]{bec18}.  \citet{freese10}, \citet{z10b}, \citet{z10a} and \citet{if12} modeled the spectra of hot, blue Pop III dark stars. They found that these objects could be observed today even by $8-10\,\mathrm{m}$ telescopes on the ground, primarily because of their high surface temperatures (20,000 - 30,000 K), larger masses (up to 10$^7$ \Ms) and longer lives \citep[up to 10$^7$ yr; see also recent reviews by][]{ds16,ban19}. Most recently, \citet{sur18a} calculated spectra for cool, red SMSs and found that some will be visible to the {\em James Webb Space Telescope} \citep[{\em JWST};][]{jwst,jwst2} at $z \lesssim$ 20 and at $z \sim$ 10 - 12 to {\em Euclid} \citep{euclid} and the {\em Wide-Field Infrared Space Telescope} \citep[{\em WFIRST};][]{wfirst} if they are gravitationally lensed. \citet{til18a} also found that the relics of such stars would be uniquely identifiable with the gravitational wave detector LISA at $z >$ 15 if they form in binaries. 

There are two challenges to modeling spectra for blue SMSs.  First, unlike the cool, red stars in \citet{sur18a}, blue SMSs cannot be approximated as blackbodies (BBs) because they have much higher ionising fluxes due to their higher surface temperatures, and much of this flux is absorbed by their own atmospheres. Second, these stars are deeply embedded in hot, dense, accretion shrouds that reprocess flux from the star into longer wavelengths. Accurate spectra for blue SMSs require both stellar atmosphere models and radiative transfer through the envelope of the star. Such spectra are crucial to predicting detections of blue SMSs at high redshift, which would capture primordial quasars at the earliest stages of their development. Here, we calculate spectra and NIR magnitudes for hot, blue Pop III SMSs at high redshift with the TLUSTY and Cloudy codes. Our models are described in Section~2, and we discuss spectra, NIR magnitudes and detection rates for blue SMSs in Section~3. We conclude in Section~4.

\begin{figure*} 
\begin{center}
\begin{tabular}{cc}
\epsfig{file=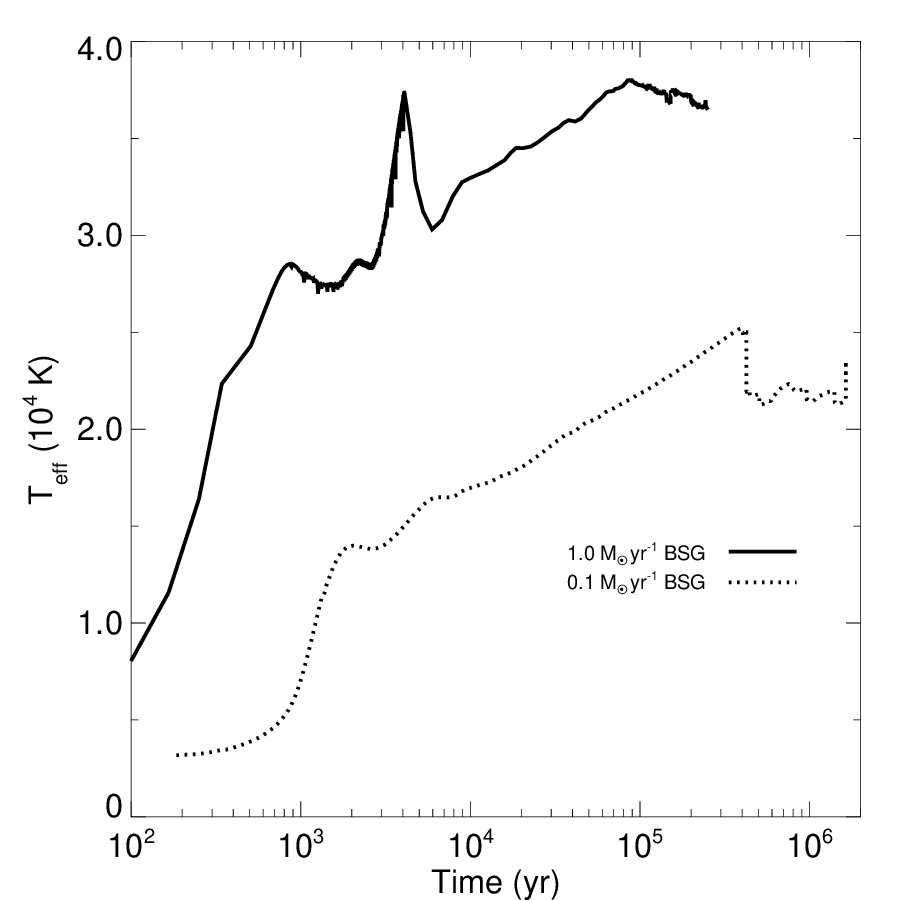,width=0.41\linewidth,clip=}  &
\epsfig{file=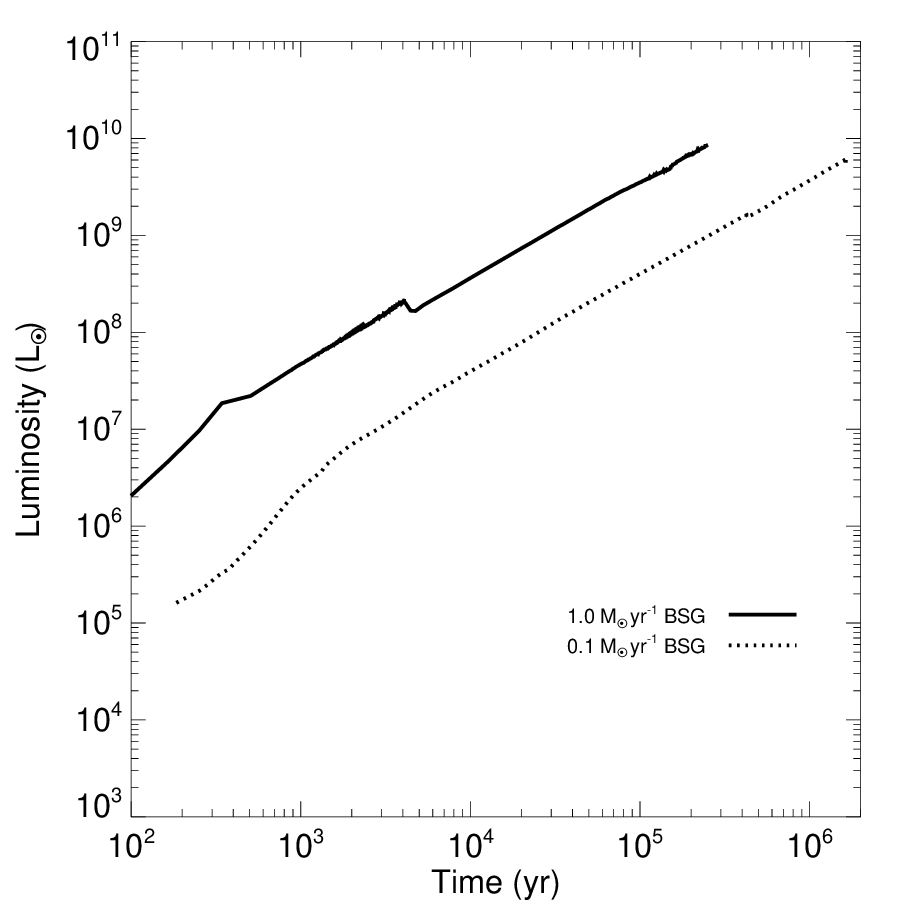,width=0.41\linewidth,clip=}  \\
\epsfig{file=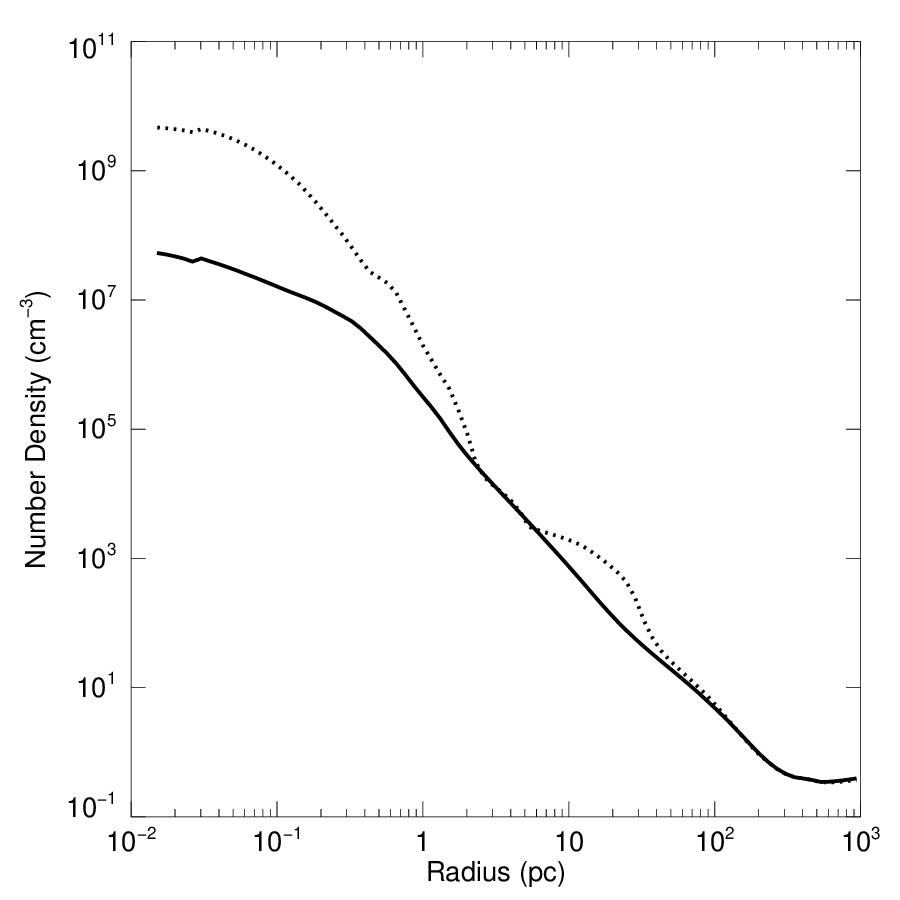,width=0.41\linewidth,clip=}  &
\epsfig{file=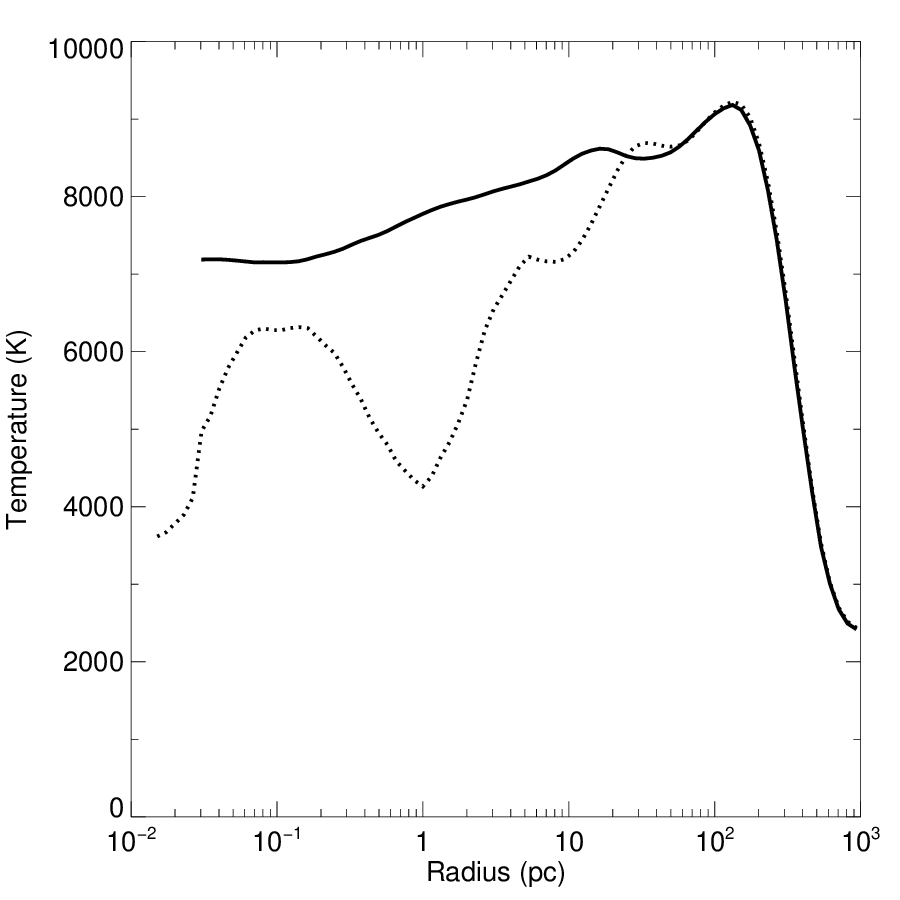,width=0.41\linewidth,clip=}  
\end{tabular}
\end{center}
\caption{Top row: evolution of SMSs accreting at 1.0 and 0.1 \Ms\ yr$^{-1}$ in our Kepler models. Left panel: surface temperatures. Right panel: luminosities. Evolution lines for the two stars end at different times because the less rapidly accreting star lives for a much longer time. Bottom row: spherically-averaged profiles of the dense, atomically-cooled shroud surrounding the star 0.238 Myr after the formation of the accretion disk. Left: gas densities. Right: temperatures.}
\label{fig:evol} 
\end{figure*}

\section{Numerical Method}

We model the atmospheres and spectra of blue SMSs with the TLUSTY code \citep{tlusty} and how their accretion envelopes reprocess these spectra with the Cloudy code \citep{cloudy17}. The emergent spectra are then cosmologically redshifted and convolved with filter functions to obtain their NIR AB magnitudes today. 

\subsection{TLUSTY Atmosphere Models} 

We consider SMSs accreting at 1.0 \Ms\ yr$^{-1}$ and 0.1 \Ms\ yr$^{-1}$ as fiducial cases. These stars were evolved in the Kepler stellar evolution code and discussed in detail in \citet{tyr17}. Surface temperatures, $T_{\mathrm{eff}}$, and bolometric luminosities for both stars over their lifetimes are shown in the upper row of Figure~\ref{fig:evol}. The spectrum of the 1.0 \Ms\ yr$^{-1}$ star was calculated at 1.51 $\times$ 10$^5$ yr, about halfway through its lifetime of 2.51 $\times$ 10$^5$ yr, when it has a mass of 1.51 $\times$ 10$^5$ \Ms\ and a surface temperature $T_{\mathrm{eff}} =$ 36,963 K. The spectrum of the 0.1 \Ms\ yr$^{-1}$ star was calculated at 8.01 $\times$ 10$^6$ yr, about halfway through its lifetime of 1.63 $\times$ 10$^6$ yr, when it has a mass of 8.01 $\times$ 10$^4$ \Ms\ and a $T_{\mathrm{eff}} =$ 22,093 K. The bolometric luminosities of the two SMSs are 1.89 $\times$ 10$^{43}$ erg s$^{-1}$ and 1.13 $\times$ 10$^{43}$ erg s$^{-1}$, respectively. As in \citet{sur18a}, we neglect the luminosity of the accretion shock at the surface of the stars because it is negligible at the velocities and densities of the infall onto the star (at most $\sim$ 10$^4$ \Ls).

The surface gravities of these stars are $\mathrm{log}(g)\approx 3.148$ and $\mathrm{log}(g)\approx 2.203$ for the 1.0 \Ms\ yr$^{-1}$ and 0.1 \Ms\ yr$^{-1}$ SMSs, respectively. TLUSTY has great difficulties converging for surface gravities as low as these, and we have therefore settled for spectra generated using TLUSTY v.205 with $\mathrm{log}(g)=3.25$ and 2.35 (i.e., offsets by $\Delta\mathrm{log}(g)\approx 0.1$ and 0.15). The stellar atmospheres are based on non-LTE, zero metallicity and primordial abundances of H and He. The resulting TLUSTY spectra have then been rescaled to match the actual bolometric luminosites of the two stars. Comparisons to zero-metallicity models with similar temperatures (but somewhat lower surface gravities) in the publicly available TLUSTY grids of \citet{Lanz03} and \citet{Lanz07} do not reveal significant problems due to these $\mathrm{log}(g)$ discrepancies, although we cannot rule out the possibility that we are slightly underestimating the ionizing flux in the case of the 0.1 \Ms\ yr$^{-1}$ model.

\subsection{Cloudy Models}

We use the TLUSTY spectra of both stars as the input spectra in our Cloudy models of the flux that emerges from the accretion envelopes of the stars, whose spherically-averaged density and temperature profiles are shown in the bottom row of Figure~\ref{fig:evol}. They are taken from an Enzo cosmology code \citep{enzo} simulation of the collapse of an atomically cooled halo after the formation of the accretion disk that creates the star \cite[see also Figure 2 of][]{sur18a}. These envelope models do not account for feedback from the SMS perturbing the structure of the infalling gas, but it is not expected to be important because ionizing radiation from the star is trapped close to its surface, as we discuss in the next section. We surround the 1.0 \Ms\ yr$^{-1}$ SMS with the profile at 0.238 Myr after the formation of the disk and the 0.1 \Ms\ yr$^{-1}$ SMS with the profile at 1.738 Myr because the envelope has time to build up to higher central densities with the more slowly accreting star. These profiles are tabulated in Cloudy with 70 bins that are uniformly partitioned in log radius, with inner and outer boundaries at 0.015 pc and 927 pc. The temperatures in the envelope are set by the virialization of cosmic flows well above it rather than by radiation from the star because ionising UV from the star is confined to very small radii deep in the cloud. Since these temperatures determine to what degree the envelope is collisionally excited, and therefore how it reprocesses photons from the star, we require Cloudy to use the Enzo temperatures for the envelope instead of inferring them from the spectrum of the star.

\begin{figure*} 
\begin{center}
\begin{tabular}{cc}
\epsfig{file=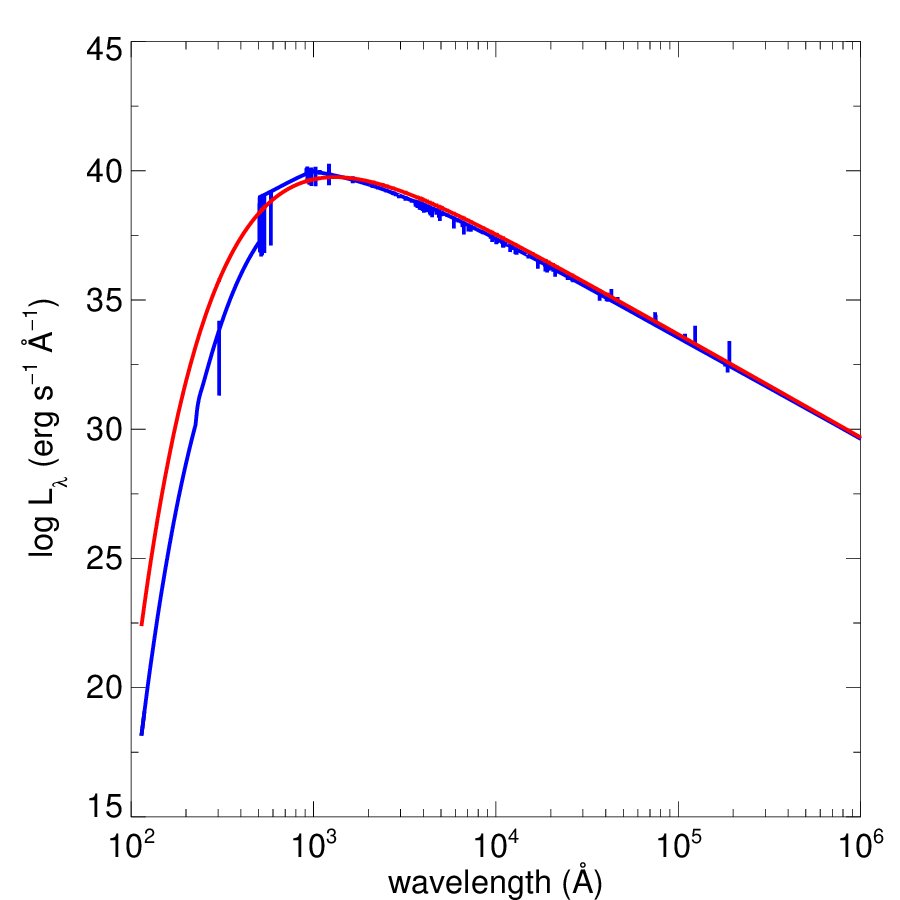,width=0.41\linewidth,clip=}  &
\epsfig{file=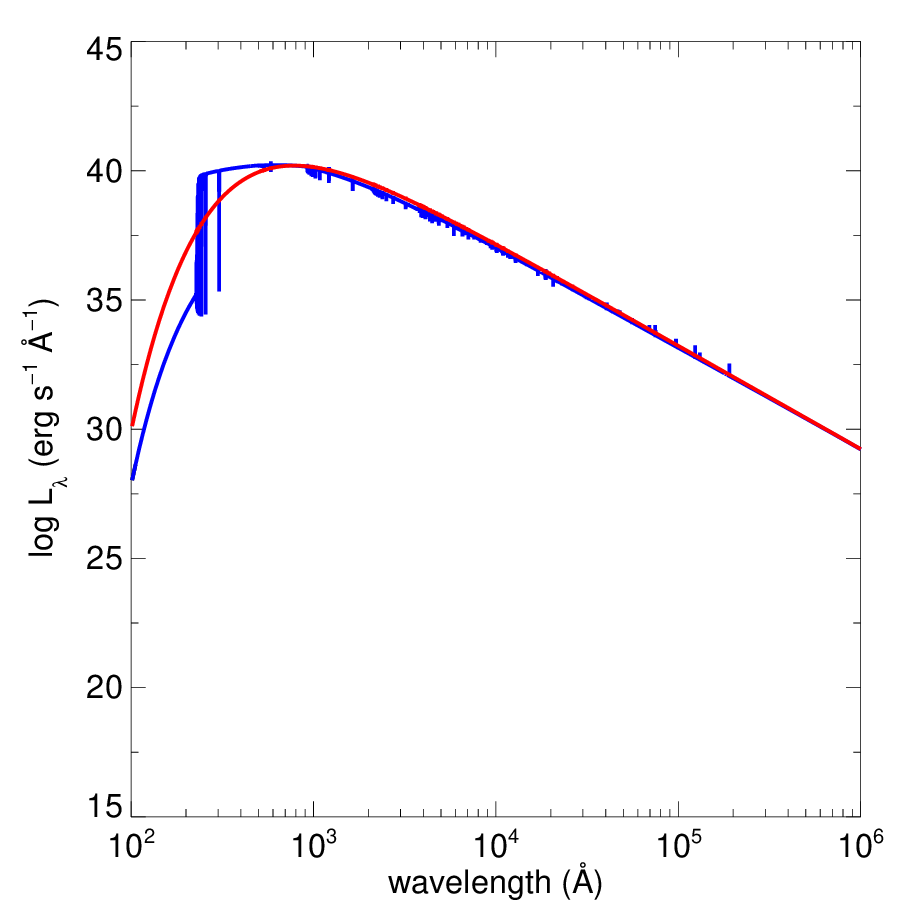,width=0.41\linewidth,clip=}  \\
\epsfig{file=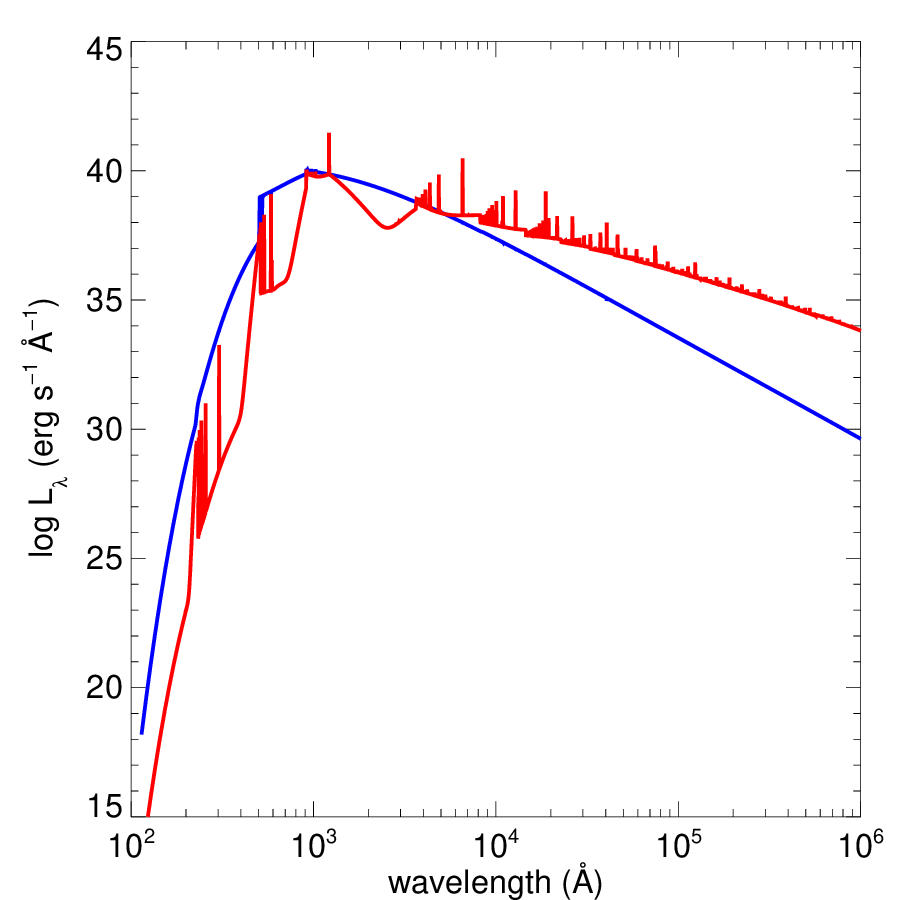,width=0.41\linewidth,clip=}  &
\epsfig{file=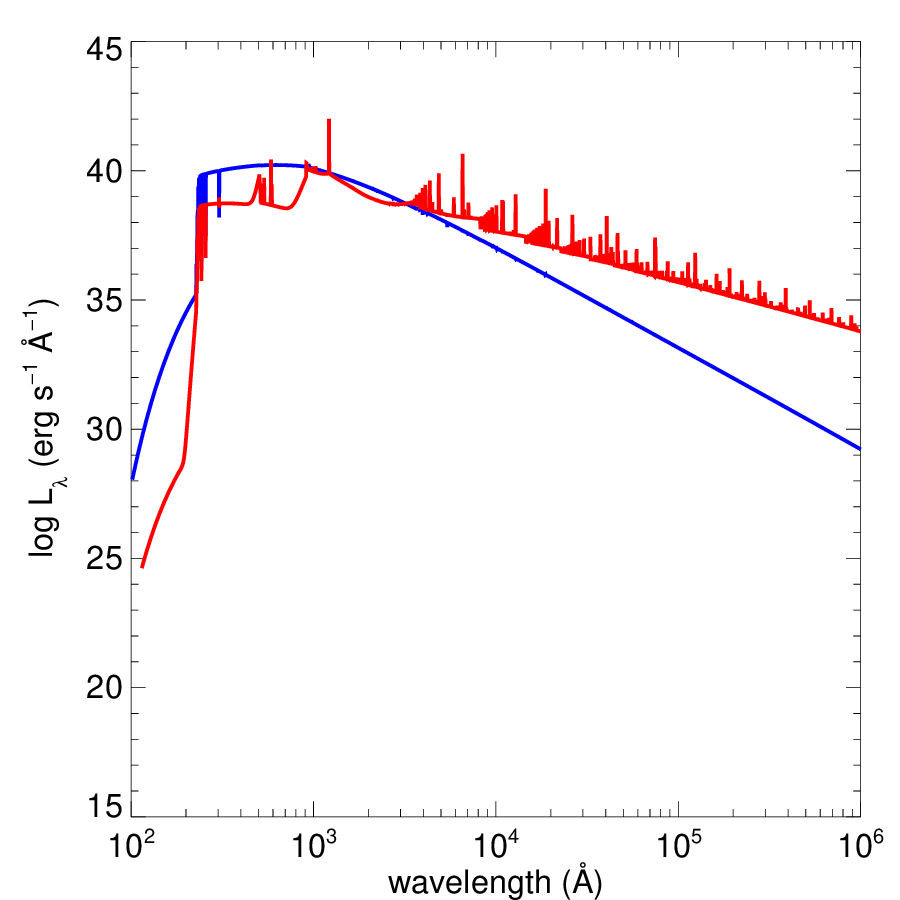,width=0.41\linewidth,clip=}  
\end{tabular}
\end{center}
\caption{Top row: spectra of the two blue SMSs in our study. Red: uncorrected BB spectrum. Blue: TLUSTY model. Left: the 0.1 \Ms\ yr$^{-1}$ star at 8.01 $\times$ 10$^4$ yr and $T_{\mathrm{eff}} = $ 22,093 K. Right: the 1.0 \Ms\ yr$^{-1}$ star at 1.08 $\times$ 10$^5$ yr and $T_{\mathrm{eff}} = $ 36,963 K. Bottom row: spectra emerging from the accretion envelopes of the stars. Blue: incident stellar spectrum. Red: spectrum after reprocessing by the envelope. Left: the 0.1 \Ms\ yr$^{-1}$ star. Right: the 1.0 \Ms\ yr$^{-1}$ star.}
\vspace{0.1in}
\label{fig:bspec} 
\end{figure*}

Cloudy then solves the equations of radiative transfer, statistical and thermal equilibrium, ionisation  and recombination, and heating and cooling to determine the excitation and ionisation state of the gas surrounding the star and calculate its emergent spectrum. These calculations use tables of recombination coefficients from \citet{dere97} and \citet{landi12} and ionic emission data from \citet{badn03} and \citet{badn06}. Each spectrum has 8228 bins that are uniformly partitioned in log $\lambda$.  We convert the luminosity in each bin, $L(\lambda) = \lambda L_{\lambda}$ in erg s$^{-1}$, to the flux density, $F_{\lambda}$ in erg s$^{-1}$ cm$^{-2}$ $\mu$m$^{-1}$, needed to compute AB magnitudes \citep[equations 1 - 3 in][]{ryd18a} by
\begin{equation}
F(\lambda) = \frac{L\left(\frac{\lambda}{1+z}\right)}{\frac{\lambda}{1+z} (1+z) 4 \pi d^2_{\mathrm{L}}(z)}.
\end{equation}
Here, $\lambda$ is the wavelength in the observer frame and $d_{\mathrm{L}}(z)$ is the luminosity distance: 
\begin{equation}
d_{\mathrm{L}}(z) = (1+z) c/\mathrm{H}_0 \int_0^z \frac{1}{\sqrt{\Omega_{\mathrm{M}} (1+z)^3 + \Omega_{\lambda}}} dz.
\end{equation}
This is done to conform to the Cloudy convention that
\begin{equation}
\sum_{\lambda}^{} \frac{L(\lambda)}{\lambda} \Delta \lambda = L_{\mathrm{bol}}.
\end{equation}
AB magnitudes, $m_{\mathrm{AB}}$, in specific filters are then calculated from
\begin{equation}
m_{\mathrm{AB}} = -2.5 \, \mathrm{log_{10}} \frac{\int_0^{\infty} F(\lambda) T(\lambda) d \lambda}{\int_0^{\infty} F_0(\lambda) T(\lambda) d \lambda}.
\end{equation}
Here, $T(\lambda)$ is the filter transmission function and $F_0(\lambda) = 3.630781 \times 10^{-20} c \lambda^{-2}~\mathrm{ergs}~\mathrm{cm}^{-2}~\mathrm{s}^{-1}~\mathrm{\mu m}^{-1}$, the reference spectrum for AB magnitudes. We assume cosmological parameters from the second-year \textit{Planck} release: $\Omega_{\mathrm{M}}=0.308$, $\Omega_{\Lambda}=0.691$, $\Omega_{\mathrm{b}}=0.0223$, $h=0.677$, $\sigma_8=0.816$, and $n=0.968$ \citep{planck2}. Flux blueward of Ly$\alpha$ in the rest frame of the star is set to zero in the AB magnitude calculation because of absorption by the neutral IGM at $z >$ 6.

\section{Blue Supermassive Stars}

\begin{figure*} 
\begin{center}
\begin{tabular}{cc}
\epsfig{file=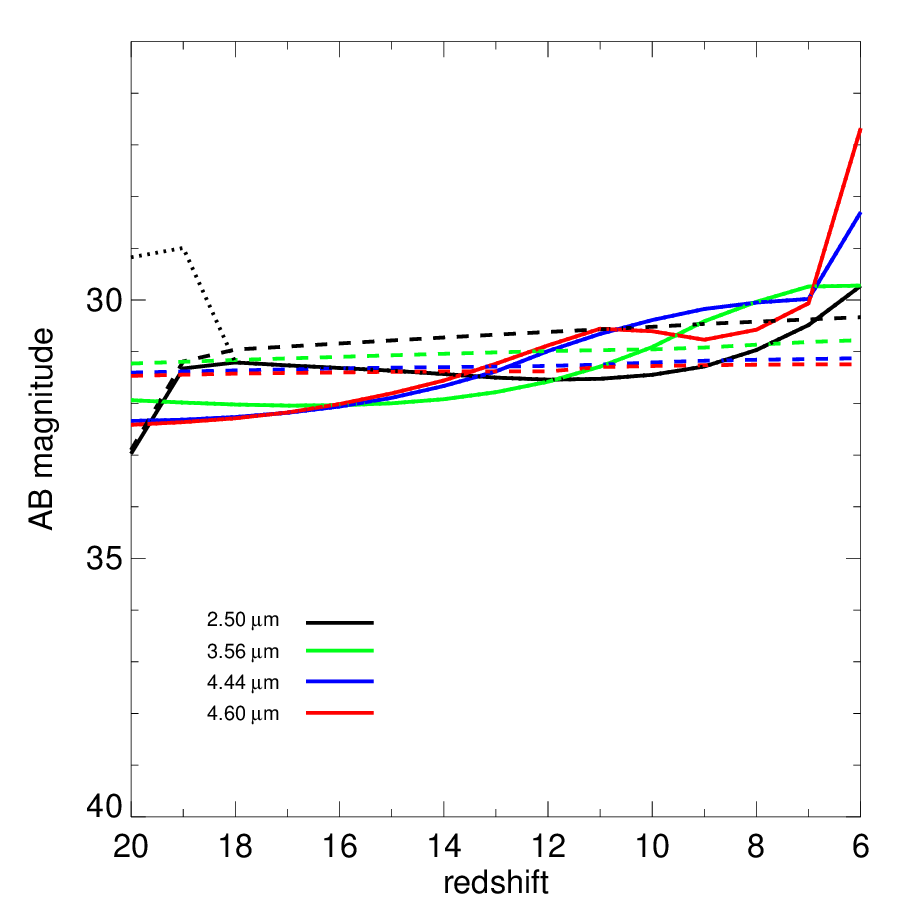,width=0.41\linewidth,clip=}  &
\epsfig{file=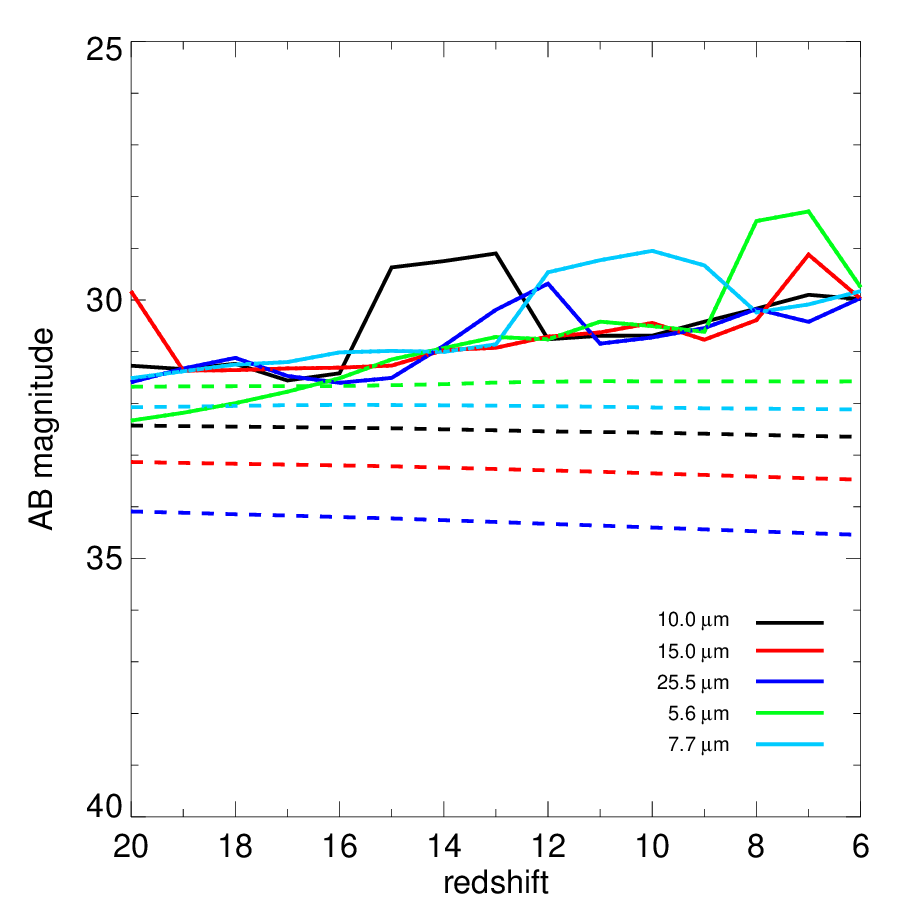,width=0.41\linewidth,clip=}  \\
\epsfig{file=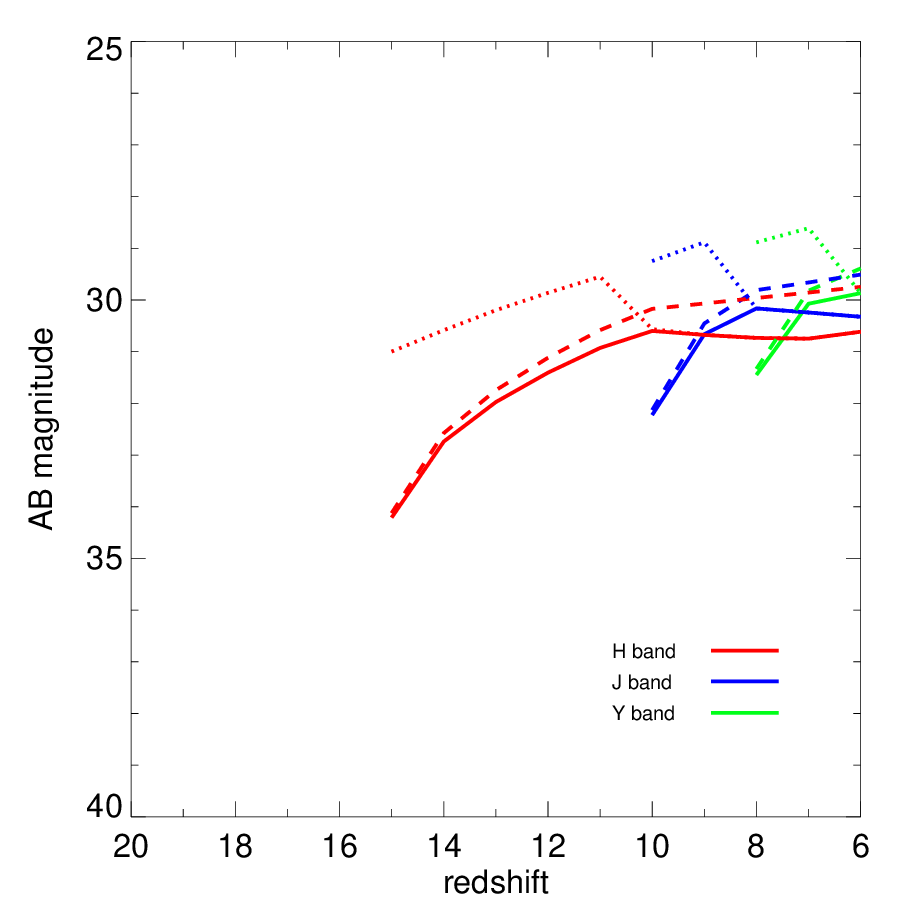,width=0.41\linewidth,clip=}  &
\epsfig{file=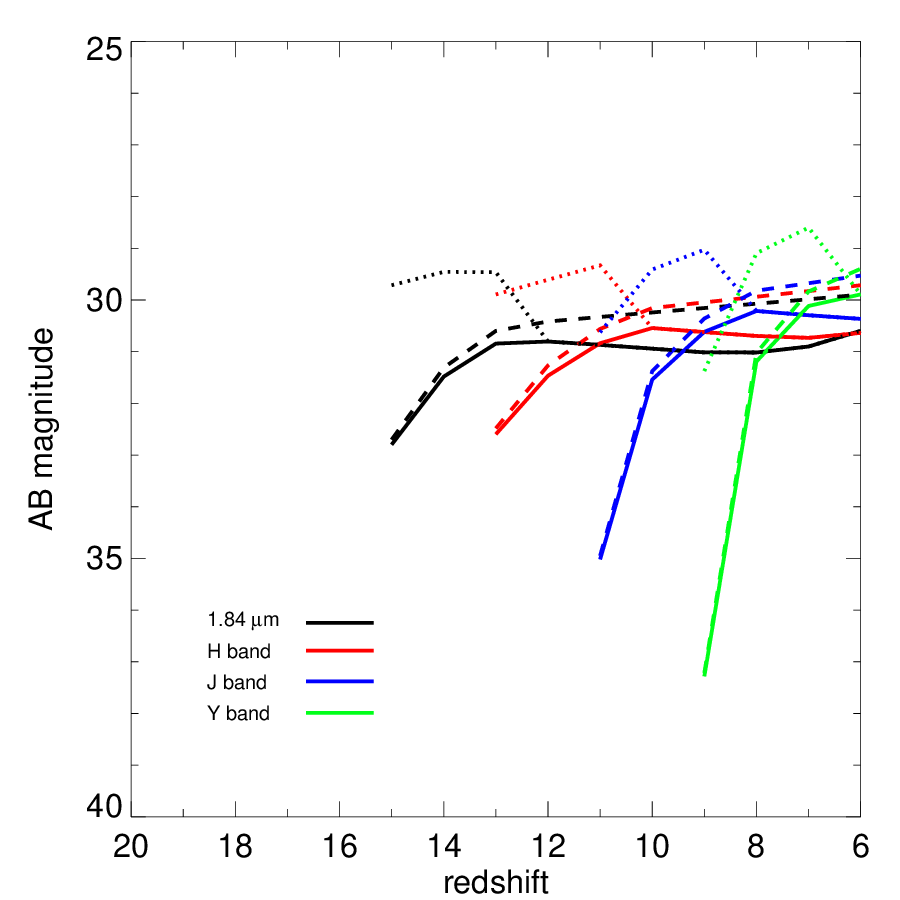,width=0.41\linewidth,clip=}  
\end{tabular}
\end{center}
\caption{NIR AB magnitudes for the 1.0 \Ms\ yr$^{-1}$ blue SMS with in {\em JWST}, {\em Euclid} and {\em WFIRST} bands. Solid line: with the accretion envelope but no contribution from its Ly$\alpha$ line. Dashed line: no envelope. Dotted line: with the envelope and its Ly$\alpha$ line. Top left: {\em JWST} NIRCam bands. Top right: {\em JWST} MIRI bands. Bottom left: {\em Euclid filters}. Bottom right: {\em WFIRST filters}.}
\vspace{0.1in}
\label{fig:1.0ABmag} 
\end{figure*}

\subsection{Stellar Spectra}

We compare TLUSTY spectra for the blue 0.1 \Ms\ yr$^{-1}$ and 1.0 \Ms\ yr$^{-1}$ stars to those of blackbodies at the same temperatures and luminosities in the upper panels of Figure~\ref{fig:bspec}. In both cases the atmosphere of the star has little effect on its spectrum at wavelengths redward of its blackbody peak except for some relatively weak emission and absorption lines, but the picture is different at shorter wavelengths. The sharp drops in luminosity at 504 \AA\ in the 0.1 \Ms\ yr$^{-1}$ star and at 227 \AA\ in the 1.0 \Ms\ yr$^{-1}$ star are due to the ionisation limits of He I and He II, respectively. There is virtually no absorption due to hydrogen just blueward of its ionisation limit except for small features at 912 \AA\ because most of it has been been ionised by the star. There is a weak Ly$\alpha$ line at 1216 \AA\ and H$\alpha$ and weak Paschen lines at 6560 \AA, 12800 \AA, and 18,800 \AA. 

\begin{figure*} 
\begin{center}
\begin{tabular}{cc}
\epsfig{file=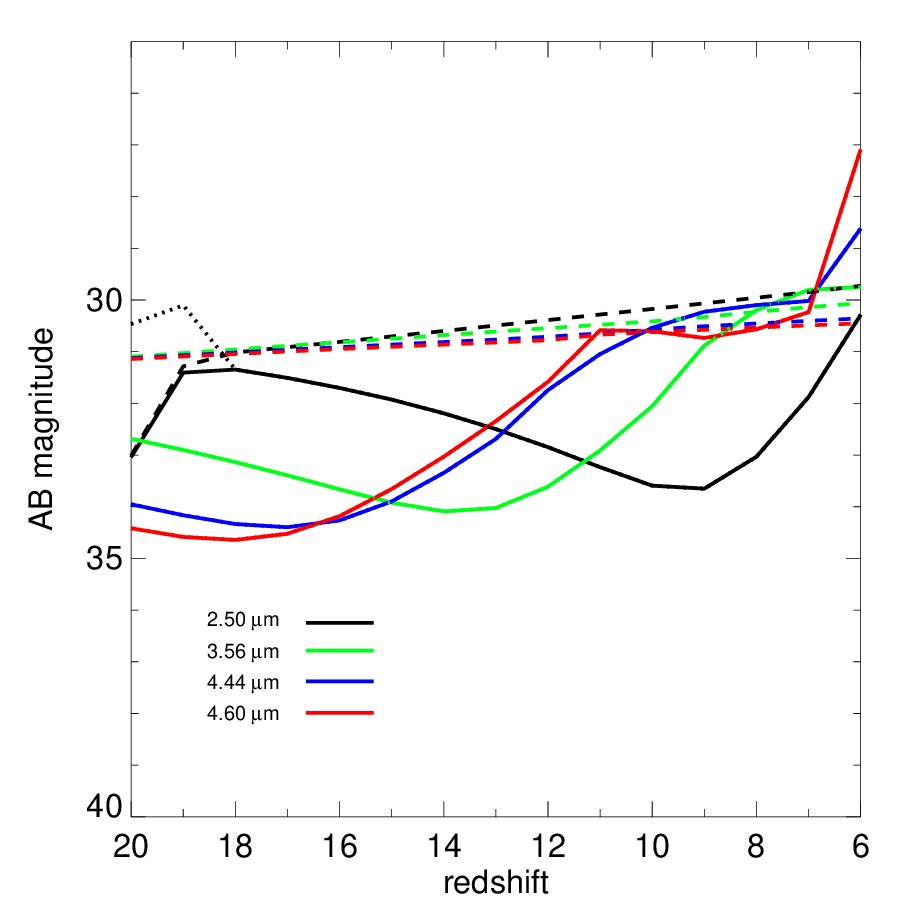,width=0.41\linewidth,clip=}  &
\epsfig{file=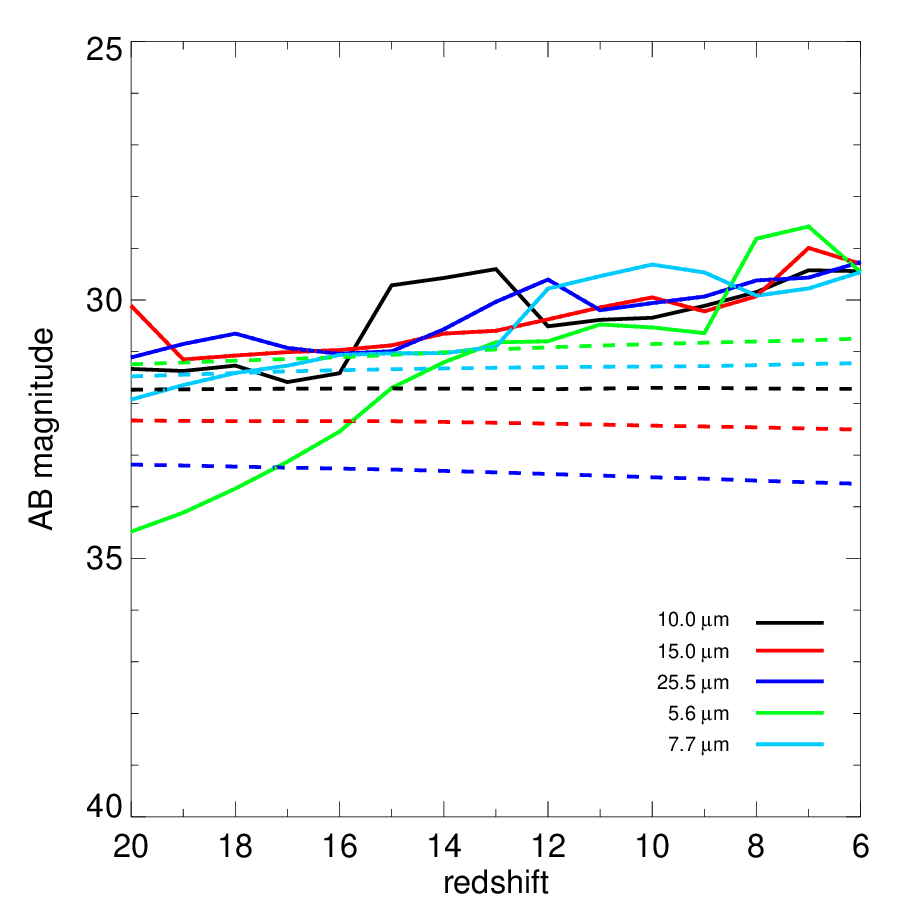,width=0.41\linewidth,clip=}  \\
\epsfig{file=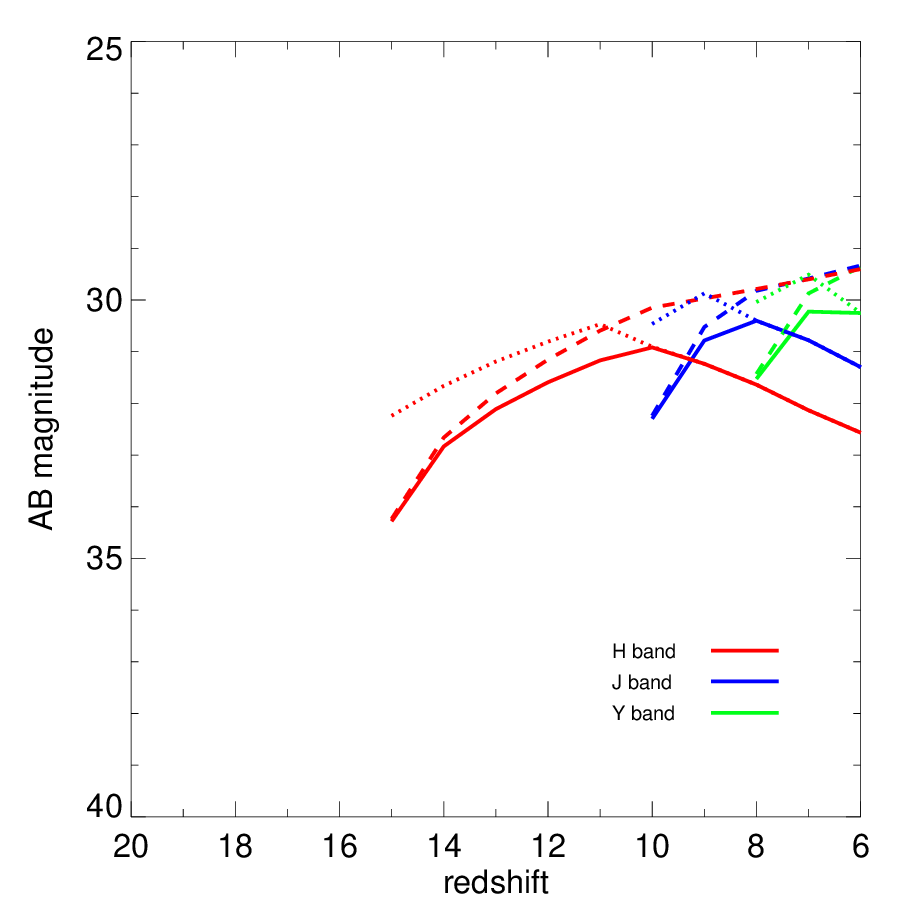,width=0.41\linewidth,clip=}  &
\epsfig{file=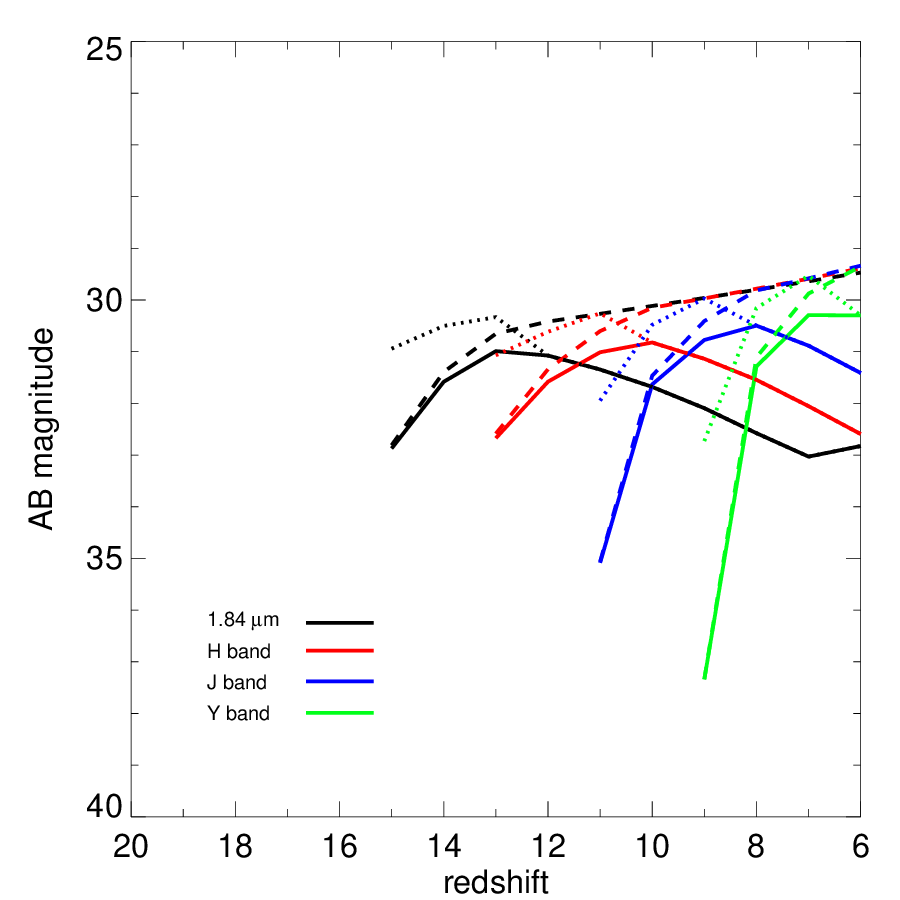,width=0.41\linewidth,clip=}  
\end{tabular}
\end{center}
\caption{NIR AB magnitudes for the 0.1 \Ms\ yr$^{-1}$ blue SMS in {\em JWST}, {\em Euclid} and {\em WFIRST} bands. Solid line: with the accretion envelope but no contribution from its Ly$\alpha$ line. Dashed line: no envelope. Dotted line: with the envelope and its Ly$\alpha$ line. Top left: {\em JWST} NIRCam bands. Top right: {\em JWST} MIRI bands. Bottom left: {\em Euclid filters}. Bottom right: {\em WFIRST filters}.}
\vspace{0.1in}
\label{fig:0.1ABmag} 
\end{figure*}

\subsection{Reprocessed Spectra}

We show Cloudy spectra for the two stars after absorption and re-emission by their accretion envelopes in the bottom row of Figure~\ref{fig:bspec}. The ionizing UV fluxes of the 0.1 and 1.0 \Ms\ yr$^{-1}$ stars are trapped at radii of 0.033 pc and 0.02 pc in our Cloudy models. These are the resolution limits of the Enzo model at 0.238 and 1.738 Myr, so as expected the strong inflows quench the ionizing UV of both stars. Strong continuum absorption due to ionisation of neutral H is evident below 912 \AA, with additional steps in absorption at 504 \AA\ and 227 \AA\ due to the ionisation of He I and He II, respectively.  These features are stronger with the 0.1 \Ms\ yr$^{-1}$ star than the \Ms\ yr$^{-1}$ star because its envelope has collapsed to higher central densities by 1.786 Myr. Strong H$\alpha$ and Paschen absorption lines are visible at 6560 \AA, 12800 \AA, and 18800 \AA. Most absorption blueward of the Lyman limit is re-emitted as the continuum and numerous lines at wavelengths above 5000 \AA.  

In contrast to the red stars in \citet{sur18a}, absorption and re-emission by the accretion envelopes of blue stars do not enhance their spectra in most of the bands that would be redshifted into the NIR today. A potential exception is Ly$\alpha$: in contrast to red SMSs, both spectra exhibit very strong Ly$\alpha$ emission lines that are pumped by the much higher UV fluxes of the blue stars. Although these lines are strong it is not clear how much of this Ly$\alpha$ flux would be observed in the NIR today, for two reasons. First, many of the Ly$\alpha$ photons would be resonantly scattered into a halo of large radius but low surface brightness, so the star might not appear to be a strong point source of this flux. Second, repeated resonant scatterings broaden the line over time so some of the flux in principle could fall outside a given filter after being redshifted into the NIR today. This is not expected to be a large effect because the Ly$\alpha$ photons are only scattered at most $\sim$ 3\% from line center before their optical depth in the wings falls below unity and they stream freely through the universe \citep{aaron15}. If the maximum displacement of the photon from line center is 0.03$\lambda_0$ $=$ 36 \AA\ in the rest frame it would be $\sim$ 0.04 $\mu$m for a $z =$ 10 SMS, or about an order of magnitude smaller than the typical width of {\em JWST} wide band NIR filters. A detailed treatment of Ly$\alpha$ radiative transfer in the primordial IGM is beyond the scope of this paper so we calculate AB magnitudes for the stars with and without the Ly$\alpha$ line as upper and lower limits.

\subsection{NIR Magnitudes}

NIR magnitudes for the 1.0 \Ms\ yr$^{-1}$ blue SMS in {\em JWST}, {\em Euclid} and {\em WFIRST} bands are plotted in Figure~\ref{fig:1.0ABmag}. We consider three cases: i) stars with accretion envelopes but no Ly$\alpha$ line; ii) stars with accretion envelopes and their Ly$\alpha$ lines; iii) stars with no envelopes. This latter case is in the event that ionising UV radiation from the star or other dynamical effects such as gravitational torqueing from nearby haloes strip away the envelope of the star.  At 2.5 - 4.6 $\mu$m the magnitudes with and without the Ly$\alpha$ line are indistinguishable out to $z = $18, when it begins to be redshifted into the 2.5 $\mu$m {\em JWST} NIRCam filter, leading to an increase in brightness of about two  magnitudes. A similar effect is visible in this filter with the 0.1 \Ms\ yr$^{-1}$ star in Figure~\ref{fig:0.1ABmag} but is less prominent because the Ly$\alpha$ line from its envelope is weaker. 

From $z \sim$ 14 - 20 the envelope of the 1.0 \Ms\ yr$^{-1}$ star somewhat suppresses flux from the star but enhances it at $z >$ 13, especially at $z <$ 7 where reprocessed radiation redward of 5000 \AA\ enhances brightnesses by 3 - 4 magnitudes. Similar enhancements are evident with the 0.1 \Ms\ yr$^{-1}$ star at the same redshifts. Absorption in the NIR by the denser envelope of the 0.1 \Ms\ yr$^{-1}$ star is more severe, decreasing its brightness down to $z \sim$ 7. In contrast, reprocessing of the spectra redward of 5000 \AA\ by the envelopes of both stars makes them more visible at 7.7 - 25.5 $\mu$m at nearly all redshifts, but their magnitudes remain well below MIRI detection limits. NIRCam AB magnitude limits of $\sim$ 31 will effectively limit detections of the 1.0 and 0.1 \Ms\ yr$^{-1}$ stars by {\em JWST} to $z \sim$ 12 and 10, respectively. 

The 1.0 \Ms\ yr $^{-1}$ star is brighter by $\sim$ 2 magnitudes in the {\em Euclid} and {\em WFIRST} filters with the Ly$\alpha$ line than without it at redshifts over which it is shifted into these filters, as shown in the lower panels of Figure~\ref{fig:1.0ABmag}. The 0.1 \Ms\ yr$^{-1}$ star is about one magnitude brighter. Exclusion of this line results in brightnesses that are consistently lower than those for stars without envelopes, and this effect is especially pronounced at lower redshifts where quenching by the envelope is greatest. Quenching at low redshifts is greatest with the 0.1 \Ms\ yr$^{-1}$ star because it has the denser envelope. The AB magnitudes of stars with envelopes but no Ly$\alpha$ never rise above 30 and could not be directly detected at $z \gtrsim$ 6 with {\em Euclid} or {\em WFIRST}, whose practical detection limits are 26 and 28, respectively.

\subsection{SMS Detection Rates}

The number of SMSs per unit redshift per unit solid angle at a redshift $z$ is
\begin{equation}
\frac{{\rm d}N}{{\rm d}z{\rm d}\Omega} = \dot{n}_{\rm SMS} \, t_{\rm SMS} \, r^{2} \frac{{\rm d}r}{{\rm d}z},
\end{equation}
where $\dot{n}_{\rm SMS}$ is the SMS formation rate per unit comoving volume, $t_{\rm SMS}$ is the average lifetime of an SMS, which we take to be 1 Myr, and $r(z)$ is the comoving distance to redshift $z$, 
\begin{equation}
r(z) = \frac{c}{H_{0}} \int_{0}^{z} \frac{{\rm d}z^{\prime}}{\sqrt{\Omega_{\rm m} (1 + z^{\prime})^{3} + \Omega_{\Lambda}}}.
\end{equation}
Current estimates of $\dot{n}_{\rm SMS}$ vary by up to eight orders of magnitude \citep{titans}, and these models also predict a variety of evolution of $\dot{n}_{\rm SMS}$ with redshift. \citet{hab16} find that the comoving number density of SMSs rises with decreasing $z$ whereas \citet{rosa17} predict that most SMSs form in the narrow range $z \sim$ 16 - 18. 

As in \citet{sur18a} we consider upper and lower limits for $\dot{n}_{\rm SMS}$. The upper limit is the low $J_{\rm crit}$ model of \citet{hab16}, in which most SMSs form at $z \sim$ 10 - 12 and the final comoving SMS number density is $\sim$ $10^{-1} \, {\rm Mpc^{-3}}$. The lower limit is found by assuming that most SMSs form at $z \sim$ 16 - 18, as in \citet{rosa17}, with a final comoving number density of $\sim$ $10^{-8} \, {\rm Mpc^{-3}}$. The upper limit yields about $4 \times 10^{7}$ SMSs per steradian per unit redshift, or around 30 per NIRCam field of view. The lower limit on $\dot{n}_{\rm SMS}$ yields only $\sim 10$ SMSs per steradian per unit redshift, or at most $10^{-5}$ SMSs per NIRCam field of view. There is also some uncertainty in SMS detection rates due to their range of lifetimes, but it is small compared to the uncertainty in $\dot{n}_{\rm SMS}$.

At present, the relative numbers of blue and red SMSs are not known. Although most studies so far have found rapidly accreting SMSs to have extended red envelopes, the codes used to model their evolution lack detailed radiation hydrodynamics and opacities and can only approximate convective mixing, all of which can have profound effects on the structure of the star. Neither blue nor red SMSs have been found in the {\em Hubble} Ultra Deep Field to date because its AB mag limit is 29 at 1.38 $\mu$m, well below that expected of either type of star even at $z \sim$ 6. Strategies for the direct detection of SMSs by {\em JWST}, {\em Euclid} and {\em WFIRST} are now under development \citep{til19a}.

\section{Discussion and Conclusion}

In contrast to cooler, redder SMSs that can be found at $z \sim$ 18 - 20, detections of hot, blue SMSs by {\em JWST} will be limited to $z \lesssim$ 10 - 12 due to quenching by their accretion envelopes. Likewise, these stars cannot be directly detected by {\em Euclid} or {\em WFIRST} at $z \gtrsim$ 6. This does not mean that these two missions cannot find blue SMSs because only moderate gravitational lensing is required to boost their fluxes above their detection limits.  Their fields of view will enclose thousands of massive galaxies and galaxy clusters, and at $z \sim$ 6 - 12 magnification factors of only 10 - 100 would be required to reveal either star.  It is likely that a significant fraction of their wide fields will be magnified by such factors \citep[e.g.,][]{om10,ryd18a}.  Higher magnifications may be possible in future surveys of individual cluster lenses by {\em JWST} but at the cost of smaller lensing volumes \citep{wet13c,wind18} that enclose fewer objects.

How could blue rapidly accreting SMSs be distinguished from hot blue dark stars of similar mass and redshift? Perhaps the greatest distinction between the two objects is the dense accretion shroud of the SMS, which imprints prominent continuum absoption features on the its spectra redward of Ly$\alpha$ in the rest frame that are absent from those of blue dark stars \citep[compare Figures 2c and 2d to Figure 6 in][]{ds16}.  In principle, these spectral features could be used to distinguish blue SMSs from hot dark stars of similar mass. Blue SMS spectra also exhibit very prominent Ly$\alpha$ lines due to pumping of the accretion envelope by high-energy UV photons from the star. Dark star spectra lack this feature because they do not have dense envelopes but, as discussed earlier, it is not clear if it could be detected today because of resonant scattering of Ly$\alpha$ by the neutral IGM at $z > $ 6.

The photospheric temperatures of supermassive Pop III stars in atomically cooling haloes (whether they are red, blue, or both in the rest frame) remain an unsolved problem. Although numerical simulations to date broadly agree on the evolution and final fates of rapidly-accreting Pop III SMSs there are key differences between them that remain poorly understood, such as the growth of the convective core mass, the final masses of the most rapidly accreting objects, and the inflation of the photosphere.  But there is a general consensus that these discrepancies likely arise from differences in how the models flag the onset of convection (i.e., the Schwarzschild or Ledoux criteria), their ability to follow dynamical instabilities (e.g., KEPLER) or not \citep[e.g., GENEVA and][]{yb08}, and their numerical resolution and boundary conditions at the surface.

Using a code derived from \citet{yb08}, \citet{hos13} found that H$^{-}$ opacity in the outermost envelopes of SMSs can greatly expand their photospheres and limit their temperatures to $\sim 0.5$--$1\times 10^{4}$K until becoming blue at masses $\gtrsim 10^{5}M_{\odot}$. \citet{tyr17}, however, find SMSs to be compact and blue from early times in the KEPLER stellar evolution code, without enough H$^{-}$ in their atmospheres to expand and cool them. A third study by \citet{hle17} with the GENEVA code found that rapidly-accreting SMSs are persistently red throughout their lifetimes, although more slowly-accreting ones may be blue (see discussion below). Efforts to benchmark these studies and converge on a solution continue \citep[see, e.g., the recent review by][]{titans}, but the final answer may only come from observations like those proposed here.

Our simulations neglect the effect of radiation pressure due to flux from the star on the flows that create it. Including these effects in cosmological simulations is challenging because they must resolve photospheres, the inner regions of accretion disks, and how the two are connected on very small scales that prevent codes from evolving them for long times. \citet{aaron17} post processed simulations of highly-resolved atomically cooling haloes with Ly$\alpha$ photon transport and found it could alter flows onto the star. Radiation hydrodynamical calculations by \citet{Luo18} and \citet{ard18} without resonant Ly$\alpha$ scattering found that radiation from the protostar did not significantly alter local flows at early times but did suppress fragmentation, thus promoting the rapid growth of a single, supermassive object.  There is somewhat more of a possibility that ionising flux from blue SMSs could blow out gas from the disk and partially expose it to the IGM, but this will simply result in AB magnitudes closer to those of the bare star shown in Figures~\ref{fig:1.0ABmag} and \ref{fig:0.1ABmag}.

Pulsations in blue or red SMSs \citep[e.g.,][]{os66} could improve their prospects for detection by temporarily boosting their fluxes above the detection limits of surveys. This phenomenon is not captured by the stellar evolution codes used here because their implicit solvers and large time steps do not resolve such oscillations, but it might cause the star to periodically brighten and dim by an order of magnitude on timescales of a few weeks to months in the rest frame. These oscillations might also facilitate their detection because their regularity would differentiate them from dusty, red high-$z$ quasars or low-$z$ interlopers such as cool Milky Way stars. Periodic dimming and brightening in principle could flag high-$z$ SMSs in transient surveys proposed for {\em JWST} such as FLARE \citep{flare}. 

The original studies on the collapse of pristine, atomically cooled haloes and the subsequent formation of SMSs assumed very high LW backgrounds that completely suppressed H$_2$ formation in their cores, so collapse was nearly isothermal at temperatures of $\sim$ 8000 K and flow rates of 0.1 - 1 \Ms\ yr$^{-1}$.  This is why we adopted them as the two fiducial rates in our study \citep[they are also typical of the collapse of atomically cooled haloes due to supersonic baryon streaming motions;][]{hir17}.  But in lower LW backgrounds some H$_2$ can form in the core of the halo and enhance cooling there, leading to lower infall rates of a few 10$^{-3}$ \Ms\ yr$^{-1}$ \citep[e.g.,][]{latif15a,rd18}.  Such rates result in much less massive stars, perhaps 10$^3$ - 10$^4$ \Ms\ rather than 10$^5$ \Ms.  It is not clear at this point which of these two populations of SMS was more prevalent in the early universe because average LW background strengths are not well understood and supersonic streaming motions are thought to have produced about as many SMSs as LW backgrounds.  Furthermore, it is not clear if these stars evolved along hot blue tracks or cool red ones, although there are indications that some would be blue \citep{hle17}. The prospects for detection of this second, less massive population of SMSs are unclear because it is not yet known if they were red or blue and they evolved in accretion envelopes with lower densities than those considered here.  But they may be more difficult to find because of their lower fluxes.  They will be studied in future work.

DCBH birth after the collapse of the SMS is the next stage of primordial quasar evolution, and a number of studies have examined prospects for their detection in future NIR surveys. These objects are also deeply embedded in dense, hot flows and techniques similar to those used here are required to model their spectra. One-dimensional radiation hydrodynamics simulations of DCBH emission post processed with Cloudy have shown that they could be detected by {\em JWST} out to $z \sim$ 20 \citep{pac15,nat17}. We will next post process radiation hydrodynamical simulations of the H II regions of DCBHs from $z =$ 10 - 20 with Cloudy to find out to what redshifts they could be found by {\em Euclid, WFIRST and JWST}.

\section*{Acknowledgements}

The authors thank the anonymous referee, whose comments improved the quality of this paper. DJW was supported by the STFC New Applicant Grant ST/P000509/1. EZ acknowledges funding from the Swedish National Space Board.  TH was supported by JSPS KAKENHI Grant Number 17F17320 and SCOG was funded by the European Research Council via the ERC Advanced Grant STARLIGHT: Formation of the First Stars (project number 339177) and from the DFG via SFB 881 ``The Milky Way System" (sub-projects B1, B2, B8). AH was supported by TDLI though a grant from the Science and Technology Commission of Shanghai Municipality (Grants No.16DZ2260200) and National Natural Science Foundation of China (Grants No.11655002). SCOG was funded by the European Research Council via the ERC Advanced Grant STARLIGHT: Formation of the First Stars (project number 339177) and from the DFG via SFB 881 ``The Milky Way System" (sub-projects B1, B2, B8). The simulations were performed on the Sciama High Performance Compute (HPC) cluster which is supported by the Institute of Cosmology and Gravitation (ICG), SEPnet and the University of Portsmouth.




\bibliographystyle{mnras}
\bibliography{refs} 

\begin{thebibliography}{}
\makeatletter
\relax
\def\mn@urlcharsother{\let\do\@makeother \do\$\do\&\do\#\do\^\do\_\do\%\do\~}
\def\mn@doi{\begingroup\mn@urlcharsother \@ifnextchar [ {\mn@doi@}
  {\mn@doi@[]}}
\def\mn@doi@[#1]#2{\def\@tempa{#1}\ifx\@tempa\@empty \href
  {http://dx.doi.org/#2} {doi:#2}\else \href {http://dx.doi.org/#2} {#1}\fi
  \endgroup}
\def\mn@eprint#1#2{\mn@eprint@#1:#2::\@nil}
\def\mn@eprint@arXiv#1{\href {http://arxiv.org/abs/#1} {{\tt arXiv:#1}}}
\def\mn@eprint@dblp#1{\href {http://dblp.uni-trier.de/rec/bibtex/#1.xml}
  {dblp:#1}}
\def\mn@eprint@#1:#2:#3:#4\@nil{\def\@tempa {#1}\def\@tempb {#2}\def\@tempc
  {#3}\ifx \@tempc \@empty \let \@tempc \@tempb \let \@tempb \@tempa \fi \ifx
  \@tempb \@empty \def\@tempb {arXiv}\fi \@ifundefined
  {mn@eprint@\@tempb}{\@tempb:\@tempc}{\expandafter \expandafter \csname
  mn@eprint@\@tempb\endcsname \expandafter{\@tempc}}}

\bibitem[\protect\citeauthoryear{{Agarwal}, {Smith}, {Glover}, {Natarajan}  \&
  {Khochfar}}{{Agarwal} et~al.}{2016}]{agarw15}
{Agarwal} B.,  {Smith} B.,  {Glover} S.,  {Natarajan} P.,   {Khochfar} S.,
  2016, \mn@doi [\mnras] {10.1093/mnras/stw929}, \href
  {http://adsabs.harvard.edu/abs/2016MNRAS.459.4209A} {459, 4209}

\bibitem[\protect\citeauthoryear{{Alvarez}, {Wise}  \& {Abel}}{{Alvarez}
  et~al.}{2009}]{awa09}
{Alvarez} M.~A.,  {Wise} J.~H.,   {Abel} T.,  2009, \mn@doi [\apjl]
  {10.1088/0004-637X/701/2/L133}, \href
  {http://adsabs.harvard.edu/abs/2009ApJ...701L.133A} {701, L133}

\bibitem[\protect\citeauthoryear{{Appenzeller} \& {Fricke}}{{Appenzeller} \&
  {Fricke}}{1972}]{af72a}
{Appenzeller} I.,  {Fricke} K.,  1972, \aap, \href
  {http://adsabs.harvard.edu/abs/1972A%26A....18...10A} {18, 10}

\bibitem[\protect\citeauthoryear{{Ardaneh}, {Luo}, {Shlosman}, {Nagamine},
  {Wise}  \& {Begelman}}{{Ardaneh} et~al.}{2018}]{ard18}
{Ardaneh} K.,  {Luo} Y.,  {Shlosman} I.,  {Nagamine} K.,  {Wise} J.~H.,
  {Begelman} M.~C.,  2018, \mn@doi [\mnras] {10.1093/mnras/sty1657}, \href
  {http://adsabs.harvard.edu/abs/2018MNRAS.479.2277A} {479, 2277}

\bibitem[\protect\citeauthoryear{{Ba{\~n}ados} et~al.,}{{Ba{\~n}ados}
  et~al.}{2018}]{ban18}
{Ba{\~n}ados} E.,  et~al., 2018, \mn@doi [\nat] {10.1038/nature25180}, \href
  {http://adsabs.harvard.edu/abs/2018Natur.553..473B} {553, 473}

\bibitem[\protect\citeauthoryear{{Badnell}}{{Badnell}}{2006}]{badn06}
{Badnell} N.~R.,  2006, \mn@doi [\apjs] {10.1086/508465}, \href
  {http://adsabs.harvard.edu/abs/2006ApJS..167..334B} {167, 334}

\bibitem[\protect\citeauthoryear{{Badnell} et~al.,}{{Badnell}
  et~al.}{2003}]{badn03}
{Badnell} N.~R.,  et~al., 2003, \mn@doi [\aap] {10.1051/0004-6361:20030816},
  \href {http://adsabs.harvard.edu/abs/2003A%26A...406.1151B} {406, 1151}

\bibitem[\protect\citeauthoryear{{Banik}, {Tan}  \& {Monaco}}{{Banik}
  et~al.}{2019}]{ban19}
{Banik} N.,  {Tan} J.~C.,   {Monaco} P.,  2019, \mn@doi [\mnras]
  {10.1093/mnras/sty3298}, \href
  {http://adsabs.harvard.edu/abs/2019MNRAS.483.3592B} {483, 3592}

\bibitem[\protect\citeauthoryear{{Baumgarte} \& {Shapiro}}{{Baumgarte} \&
  {Shapiro}}{1999}]{baum99}
{Baumgarte} T.~W.,  {Shapiro} S.~L.,  1999, \mn@doi [\apj] {10.1086/308006},
  \href {http://adsabs.harvard.edu/abs/1999ApJ...526..941B} {526, 941}

\bibitem[\protect\citeauthoryear{{Becerra}, {Marinacci}, {Bromm}  \&
  {Hernquist}}{{Becerra} et~al.}{2018}]{bec18}
{Becerra} F.,  {Marinacci} F.,  {Bromm} V.,   {Hernquist} L.~E.,  2018, \mn@doi
  [\mnras] {10.1093/mnras/sty2210}, \href
  {http://adsabs.harvard.edu/abs/2018MNRAS.480.5029B} {480, 5029}

\bibitem[\protect\citeauthoryear{{Bryan} et~al.,}{{Bryan} et~al.}{2014}]{enzo}
{Bryan} G.~L.,  et~al., 2014, \mn@doi [\apjs] {10.1088/0067-0049/211/2/19},
  \href {http://adsabs.harvard.edu/abs/2014ApJS..211...19B} {211, 19}

\bibitem[\protect\citeauthoryear{{Butler}, {Lima}, {Baumgarte}  \&
  {Shapiro}}{{Butler} et~al.}{2018}]{but18}
{Butler} S.~P.,  {Lima} A.~R.,  {Baumgarte} T.~W.,   {Shapiro} S.~L.,  2018,
  \mn@doi [\mnras] {10.1093/mnras/sty834}, \href
  {http://adsabs.harvard.edu/abs/2018MNRAS.477.3694B} {477, 3694}

\bibitem[\protect\citeauthoryear{{Chandrasekhar}}{{Chandrasekhar}}{1964}]{chandra64}
{Chandrasekhar} S.,  1964, \mn@doi [\apj] {10.1086/147938}, \href
  {http://adsabs.harvard.edu/abs/1964ApJ...140..417C} {140, 417}

\bibitem[\protect\citeauthoryear{{Chen}, {Heger}, {Woosley}, {Almgren},
  {Whalen}  \& {Johnson}}{{Chen} et~al.}{2014}]{chen14b}
{Chen} K.-J.,  {Heger} A.,  {Woosley} S.,  {Almgren} A.,  {Whalen} D.~J.,
  {Johnson} J.~L.,  2014, \mn@doi [\apj] {10.1088/0004-637X/790/2/162}, \href
  {http://adsabs.harvard.edu/abs/2014ApJ...790..162C} {790, 162}

\bibitem[\protect\citeauthoryear{{Chon}, {Hosokawa}  \& {Yoshida}}{{Chon}
  et~al.}{2017}]{chon17b}
{Chon} S.,  {Hosokawa} T.,   {Yoshida} N.,  2017, arXiv:1711.05262, \href
  {http://adsabs.harvard.edu/abs/2017arXiv171105262C} {}

\bibitem[\protect\citeauthoryear{{Dere}, {Landi}, {Mason}, {Monsignori Fossi}
  \& {Young}}{{Dere} et~al.}{1997}]{dere97}
{Dere} K.~P.,  {Landi} E.,  {Mason} H.~E.,  {Monsignori Fossi} B.~C.,   {Young}
  P.~R.,  1997, \mn@doi [\aaps] {10.1051/aas:1997368}, \href
  {http://adsabs.harvard.edu/abs/1997A%26AS..125..149D} {125, 149}

\bibitem[\protect\citeauthoryear{{Ferland} et~al.,}{{Ferland}
  et~al.}{2017}]{cloudy17}
{Ferland} G.~J.,  et~al., 2017, {\em Rev. Mex. Astron \& Astrophys}, \href
  {http://adsabs.harvard.edu/abs/2017RMxAA..53..385F} {53, 385}

\bibitem[\protect\citeauthoryear{{Fowler}}{{Fowler}}{1964}]{fowler64}
{Fowler} W.~A.,  1964, \mn@doi [Reviews of Modern Physics]
  {10.1103/RevModPhys.36.545}, \href
  {http://adsabs.harvard.edu/abs/1964RvMP...36..545F} {36, 545}

\bibitem[\protect\citeauthoryear{{Fowler}}{{Fowler}}{1966}]{fowler66}
{Fowler} W.~A.,  1966, \mn@doi [\apj] {10.1086/148594}, \href
  {http://adsabs.harvard.edu/abs/1966ApJ...144..180F} {144, 180}

\bibitem[\protect\citeauthoryear{{Freese}, {Bodenheimer}, {Spolyar}  \&
  {Gondolo}}{{Freese} et~al.}{2008}]{freese08b}
{Freese} K.,  {Bodenheimer} P.,  {Spolyar} D.,   {Gondolo} P.,  2008, \mn@doi
  [\apjl] {10.1086/592685}, \href
  {https://ui.adsabs.harvard.edu/abs/2008ApJ...685L.101F} {685, L101}

\bibitem[\protect\citeauthoryear{{Freese}, {Ilie}, {Spolyar}, {Valluri}  \&
  {Bodenheimer}}{{Freese} et~al.}{2010}]{freese10}
{Freese} K.,  {Ilie} C.,  {Spolyar} D.,  {Valluri} M.,   {Bodenheimer} P.,
  2010, \mn@doi [\apj] {10.1088/0004-637X/716/2/1397}, \href
  {http://adsabs.harvard.edu/abs/2010ApJ...716.1397F} {716, 1397}

\bibitem[\protect\citeauthoryear{{Freese}, {Rindler-Daller}, {Spolyar}  \&
  {Valluri}}{{Freese} et~al.}{2016}]{ds16}
{Freese} K.,  {Rindler-Daller} T.,  {Spolyar} D.,   {Valluri} M.,  2016,
  \mn@doi [Reports on Progress in Physics] {10.1088/0034-4885/79/6/066902},
  \href {http://adsabs.harvard.edu/abs/2016RPPh...79f6902F} {79, 066902}

\bibitem[\protect\citeauthoryear{{Fuller}, {Woosley}  \& {Weaver}}{{Fuller}
  et~al.}{1986}]{fuller86}
{Fuller} G.~M.,  {Woosley} S.~E.,   {Weaver} T.~A.,  1986, \mn@doi [\apj]
  {10.1086/164452}, \href {http://adsabs.harvard.edu/abs/1986ApJ...307..675F}
  {307, 675}

\bibitem[\protect\citeauthoryear{{Gardner} et~al.,}{{Gardner}
  et~al.}{2006}]{jwst}
{Gardner} J.~P.,  et~al., 2006, \mn@doi [\ssr] {10.1007/s11214-006-8315-7},
  \href {http://adsabs.harvard.edu/abs/2006SSRv..123..485G} {123, 485}

\bibitem[\protect\citeauthoryear{{Habouzit}, {Volonteri}, {Latif}, {Dubois}  \&
  {Peirani}}{{Habouzit} et~al.}{2016}]{hab16}
{Habouzit} M.,  {Volonteri} M.,  {Latif} M.,  {Dubois} Y.,   {Peirani} S.,
  2016, \mn@doi [\mnras] {10.1093/mnras/stw1924}, \href
  {http://adsabs.harvard.edu/abs/2016MNRAS.463..529H} {463, 529}

\bibitem[\protect\citeauthoryear{{Haemmerl{\'e}} \& {Meynet}}{{Haemmerl{\'e}}
  \& {Meynet}}{2019}]{hle19}
{Haemmerl{\'e}} L.,  {Meynet} G.,  2019, arXiv:1903.00020, \href
  {https://ui.adsabs.harvard.edu/\#abs/2019arXiv190300020H} {p.
  arXiv:1903.00020}

\bibitem[\protect\citeauthoryear{{Haemmerl{\'e}}, {Woods}, {Klessen}, {Heger}
  \& {Whalen}}{{Haemmerl{\'e}} et~al.}{2018a}]{hle17}
{Haemmerl{\'e}} L.,  {Woods} T.~E.,  {Klessen} R.~S.,  {Heger} A.,   {Whalen}
  D.~J.,  2018a, \mn@doi [\mnras] {10.1093/mnras/stx2919}, \href
  {http://adsabs.harvard.edu/abs/2018MNRAS.474.2757H} {474, 2757}

\bibitem[\protect\citeauthoryear{{Haemmerl{\'e}}, {Woods}, {Klessen}, {Heger}
  \& {Whalen}}{{Haemmerl{\'e}} et~al.}{2018b}]{hle18}
{Haemmerl{\'e}} L.,  {Woods} T.~E.,  {Klessen} R.~S.,  {Heger} A.,   {Whalen}
  D.~J.,  2018b, \mn@doi [\apjl] {10.3847/2041-8213/aaa462}, \href
  {http://adsabs.harvard.edu/abs/2018ApJ...853L...3H} {853, L3}

\bibitem[\protect\citeauthoryear{{Hartwig}, {Agarwal}  \& {Regan}}{{Hartwig}
  et~al.}{2018}]{til18a}
{Hartwig} T.,  {Agarwal} B.,   {Regan} J.~A.,  2018, \mn@doi [\mnras]
  {10.1093/mnrasl/sly091}, \href
  {http://adsabs.harvard.edu/abs/2018MNRAS.479L..23H} {479, L23}

\bibitem[\protect\citeauthoryear{{Hirano}, {Hosokawa}, {Yoshida}  \&
  {Kuiper}}{{Hirano} et~al.}{2017}]{hir17}
{Hirano} S.,  {Hosokawa} T.,  {Yoshida} N.,   {Kuiper} R.,  2017, \mn@doi
  [Science] {10.1126/science.aai9119}, \href
  {http://adsabs.harvard.edu/abs/2017Sci...357.1375H} {357, 1375}

\bibitem[\protect\citeauthoryear{{Hosokawa}, {Yorke}, {Inayoshi}, {Omukai}  \&
  {Yoshida}}{{Hosokawa} et~al.}{2013}]{hos13}
{Hosokawa} T.,  {Yorke} H.~W.,  {Inayoshi} K.,  {Omukai} K.,   {Yoshida} N.,
  2013, \mn@doi [\apj] {10.1088/0004-637X/778/2/178}, \href
  {http://adsabs.harvard.edu/abs/2013ApJ...778..178H} {778, 178}

\bibitem[\protect\citeauthoryear{{Hubeny} \& {Lanz}}{{Hubeny} \&
  {Lanz}}{1995}]{tlusty}
{Hubeny} I.,  {Lanz} T.,  1995, \mn@doi [\apj] {10.1086/175226}, \href
  {http://adsabs.harvard.edu/abs/1995ApJ...439..875H} {439, 875}

\bibitem[\protect\citeauthoryear{{Iben}}{{Iben}}{1963}]{iben63}
{Iben} Jr. I.,  1963, \mn@doi [\apj] {10.1086/147708}, \href
  {http://adsabs.harvard.edu/abs/1963ApJ...138.1090I} {138, 1090}

\bibitem[\protect\citeauthoryear{{Ilie}, {Freese}, {Valluri}, {Iliev}  \&
  {Shapiro}}{{Ilie} et~al.}{2012}]{if12}
{Ilie} C.,  {Freese} K.,  {Valluri} M.,  {Iliev} I.~T.,   {Shapiro} P.~R.,
  2012, \mn@doi [\mnras] {10.1111/j.1365-2966.2012.20760.x}, \href
  {https://ui.adsabs.harvard.edu/abs/2012MNRAS.422.2164I} {422, 2164}

\bibitem[\protect\citeauthoryear{{Inayoshi}, {Omukai}  \& {Tasker}}{{Inayoshi}
  et~al.}{2014}]{iot14}
{Inayoshi} K.,  {Omukai} K.,   {Tasker} E.,  2014, \mn@doi [\mnras]
  {10.1093/mnrasl/slu151}, \href
  {http://adsabs.harvard.edu/abs/2014MNRAS.445L.109I} {445, L109}

\bibitem[\protect\citeauthoryear{{Inayoshi}, {Haiman}  \&
  {Ostriker}}{{Inayoshi} et~al.}{2016}]{inay16}
{Inayoshi} K.,  {Haiman} Z.,   {Ostriker} J.~P.,  2016, \mn@doi [\mnras]
  {10.1093/mnras/stw836}, \href
  {http://adsabs.harvard.edu/abs/2016MNRAS.459.3738I} {459, 3738}

\bibitem[\protect\citeauthoryear{{Johnson}, {Whalen}, {Fryer}  \&
  {Li}}{{Johnson} et~al.}{2012}]{jlj12a}
{Johnson} J.~L.,  {Whalen} D.~J.,  {Fryer} C.~L.,   {Li} H.,  2012, \mn@doi
  [\apj] {10.1088/0004-637X/750/1/66}, \href
  {http://adsabs.harvard.edu/abs/2012ApJ...750...66J} {750, 66}

\bibitem[\protect\citeauthoryear{{Johnson}, {Whalen}, {Li}  \&
  {Holz}}{{Johnson} et~al.}{2013a}]{jet13}
{Johnson} J.~L.,  {Whalen} D.~J.,  {Li} H.,   {Holz} D.~E.,  2013a, \mn@doi
  [\apj] {10.1088/0004-637X/771/2/116}, \href
  {http://adsabs.harvard.edu/abs/2013ApJ...771..116J} {771, 116}

\bibitem[\protect\citeauthoryear{{Johnson}, {Whalen}, {Even}, {Fryer}, {Heger},
  {Smidt}  \& {Chen}}{{Johnson} et~al.}{2013b}]{jet13a}
{Johnson} J.~L.,  {Whalen} D.~J.,  {Even} W.,  {Fryer} C.~L.,  {Heger} A.,
  {Smidt} J.,   {Chen} K.-J.,  2013b, \mn@doi [\apj]
  {10.1088/0004-637X/775/2/107}, \href
  {http://adsabs.harvard.edu/abs/2013ApJ...775..107J} {775, 107}

\bibitem[\protect\citeauthoryear{{Kalirai}}{{Kalirai}}{2018}]{jwst2}
{Kalirai} J.,  2018, \mn@doi [Contemporary Physics]
  {10.1080/00107514.2018.1467648}, \href
  {http://adsabs.harvard.edu/abs/2018ConPh..59..251K} {59, 251}

\bibitem[\protect\citeauthoryear{{Landi}, {Del Zanna}, {Young}, {Dere}  \&
  {Mason}}{{Landi} et~al.}{2012}]{landi12}
{Landi} E.,  {Del Zanna} G.,  {Young} P.~R.,  {Dere} K.~P.,   {Mason} H.~E.,
  2012, \mn@doi [\apj] {10.1088/0004-637X/744/2/99}, \href
  {http://adsabs.harvard.edu/abs/2012ApJ...744...99L} {744, 99}

\bibitem[\protect\citeauthoryear{{Lanz} \& {Hubeny}}{{Lanz} \&
  {Hubeny}}{2003}]{Lanz03}
{Lanz} T.,  {Hubeny} I.,  2003, \mn@doi [\apjs] {10.1086/374373}, \href
  {http://adsabs.harvard.edu/abs/2003ApJS..146..417L} {146, 417}

\bibitem[\protect\citeauthoryear{{Lanz} \& {Hubeny}}{{Lanz} \&
  {Hubeny}}{2007}]{Lanz07}
{Lanz} T.,  {Hubeny} I.,  2007, \mn@doi [\apjs] {10.1086/511270}, \href
  {http://adsabs.harvard.edu/abs/2007ApJS..169...83L} {169, 83}

\bibitem[\protect\citeauthoryear{{Latif} \& {Volonteri}}{{Latif} \&
  {Volonteri}}{2015}]{latif15b}
{Latif} M.~A.,  {Volonteri} M.,  2015, \mn@doi [\mnras]
  {10.1093/mnras/stv1337}, \href
  {http://adsabs.harvard.edu/abs/2015MNRAS.452.1026L} {452, 1026}

\bibitem[\protect\citeauthoryear{{Latif}, {Niemeyer}  \& {Schleicher}}{{Latif}
  et~al.}{2014a}]{lns14}
{Latif} M.~A.,  {Niemeyer} J.~C.,   {Schleicher} D.~R.~G.,  2014a, \mn@doi
  [\mnras] {10.1093/mnras/stu489}, \href
  {http://adsabs.harvard.edu/abs/2014MNRAS.440.2969L} {440, 2969}

\bibitem[\protect\citeauthoryear{{Latif}, {Bovino}, {Van Borm}, {Grassi},
  {Schleicher}  \& {Spaans}}{{Latif} et~al.}{2014b}]{latif14}
{Latif} M.~A.,  {Bovino} S.,  {Van Borm} C.,  {Grassi} T.,  {Schleicher}
  D.~R.~G.,   {Spaans} M.,  2014b, \mn@doi [\mnras] {10.1093/mnras/stu1230},
  \href {http://adsabs.harvard.edu/abs/2014MNRAS.443.1979L} {443, 1979}

\bibitem[\protect\citeauthoryear{{Latif}, {Bovino}, {Grassi}, {Schleicher}  \&
  {Spaans}}{{Latif} et~al.}{2015}]{latif15a}
{Latif} M.~A.,  {Bovino} S.,  {Grassi} T.,  {Schleicher} D.~R.~G.,   {Spaans}
  M.,  2015, \mn@doi [\mnras] {10.1093/mnras/stu2244}, \href
  {http://adsabs.harvard.edu/abs/2015MNRAS.446.3163L} {446, 3163}

\bibitem[\protect\citeauthoryear{{Laureijs} et~al.,}{{Laureijs}
  et~al.}{2011}]{euclid}
{Laureijs} R.,  et~al., 2011, arXiv:1110.3193, \href
  {http://adsabs.harvard.edu/abs/2011arXiv1110.3193L} {}

\bibitem[\protect\citeauthoryear{{Lodato} \& {Natarajan}}{{Lodato} \&
  {Natarajan}}{2006}]{ln06}
{Lodato} G.,  {Natarajan} P.,  2006, \mn@doi [\mnras]
  {10.1111/j.1365-2966.2006.10801.x}, \href
  {http://adsabs.harvard.edu/abs/2006MNRAS.371.1813L} {371, 1813}

\bibitem[\protect\citeauthoryear{{Luo}, {Ardaneh}, {Shlosman}, {Nagamine},
  {Wise}  \& {Begelman}}{{Luo} et~al.}{2018}]{Luo18}
{Luo} Y.,  {Ardaneh} K.,  {Shlosman} I.,  {Nagamine} K.,  {Wise} J.~H.,
  {Begelman} M.~C.,  2018, \mn@doi [\mnras] {10.1093/mnras/sty362}, \href
  {http://adsabs.harvard.edu/abs/2018MNRAS.476.3523L} {476, 3523}

\bibitem[\protect\citeauthoryear{{Madau}, {Haardt}  \& {Dotti}}{{Madau}
  et~al.}{2014}]{mhd14}
{Madau} P.,  {Haardt} F.,   {Dotti} M.,  2014, \mn@doi [\apjl]
  {10.1088/2041-8205/784/2/L38}, \href
  {http://adsabs.harvard.edu/abs/2014ApJ...784L..38M} {784, L38}

\bibitem[\protect\citeauthoryear{{Mayer} \& {Bonoli}}{{Mayer} \&
  {Bonoli}}{2019}]{may19}
{Mayer} L.,  {Bonoli} S.,  2019, \mn@doi [Reports on Progress in Physics]
  {10.1088/1361-6633/aad6a5}, \href
  {http://adsabs.harvard.edu/abs/2019RPPh...82a6901M} {82, 016901}

\bibitem[\protect\citeauthoryear{{Mayer}, {Fiacconi}, {Bonoli}, {Quinn}, {Ro{\v
  s}kar}, {Shen}  \& {Wadsley}}{{Mayer} et~al.}{2015}]{may15}
{Mayer} L.,  {Fiacconi} D.,  {Bonoli} S.,  {Quinn} T.,  {Ro{\v s}kar} R.,
  {Shen} S.,   {Wadsley} J.,  2015, \mn@doi [\apj]
  {10.1088/0004-637X/810/1/51}, \href
  {http://adsabs.harvard.edu/abs/2015ApJ...810...51M} {810, 51}

\bibitem[\protect\citeauthoryear{{Montero}, {Janka}  \& {M{\"u}ller}}{{Montero}
  et~al.}{2012}]{montero12}
{Montero} P.~J.,  {Janka} H.-T.,   {M{\"u}ller} E.,  2012, \mn@doi [\apj]
  {10.1088/0004-637X/749/1/37}, \href
  {http://adsabs.harvard.edu/abs/2012ApJ...749...37M} {749, 37}

\bibitem[\protect\citeauthoryear{{Mortlock} et~al.,}{{Mortlock}
  et~al.}{2011}]{mort11}
{Mortlock} D.~J.,  et~al., 2011, \mn@doi [\nat] {10.1038/nature10159}, \href
  {http://adsabs.harvard.edu/abs/2011Natur.474..616M} {474, 616}

\bibitem[\protect\citeauthoryear{{Natarajan}, {Pacucci}, {Ferrara}, {Agarwal},
  {Ricarte}, {Zackrisson}  \& {Cappelluti}}{{Natarajan} et~al.}{2017}]{nat17}
{Natarajan} P.,  {Pacucci} F.,  {Ferrara} A.,  {Agarwal} B.,  {Ricarte} A.,
  {Zackrisson} E.,   {Cappelluti} N.,  2017, \mn@doi [\apj]
  {10.3847/1538-4357/aa6330}, \href
  {http://adsabs.harvard.edu/abs/2017ApJ...838..117N} {838, 117}

\bibitem[\protect\citeauthoryear{{Oguri} \& {Marshall}}{{Oguri} \&
  {Marshall}}{2010}]{om10}
{Oguri} M.,  {Marshall} P.~J.,  2010, \mn@doi [\mnras]
  {10.1111/j.1365-2966.2010.16639.x}, \href
  {http://adsabs.harvard.edu/abs/2010MNRAS.405.2579O} {405, 2579}

\bibitem[\protect\citeauthoryear{{Osaki}}{{Osaki}}{1966}]{os66}
{Osaki} Y.,  1966, \pasj, \href
  {http://adsabs.harvard.edu/abs/1966PASJ...18..384O} {18, 384}

\bibitem[\protect\citeauthoryear{{Pacucci}, {Ferrara}, {Volonteri}  \&
  {Dubus}}{{Pacucci} et~al.}{2015}]{pac15}
{Pacucci} F.,  {Ferrara} A.,  {Volonteri} M.,   {Dubus} G.,  2015, \mn@doi
  [\mnras] {10.1093/mnras/stv2196}, \href
  {http://adsabs.harvard.edu/abs/2015MNRAS.454.3771P} {454, 3771}

\bibitem[\protect\citeauthoryear{{Pezzulli}, {Valiante}  \&
  {Schneider}}{{Pezzulli} et~al.}{2016}]{pez16}
{Pezzulli} E.,  {Valiante} R.,   {Schneider} R.,  2016, \mn@doi [\mnras]
  {10.1093/mnras/stw505}, \href
  {http://adsabs.harvard.edu/abs/2016MNRAS.458.3047P} {458, 3047}

\bibitem[\protect\citeauthoryear{{Planck Collaboration} et~al.,}{{Planck
  Collaboration} et~al.}{2016}]{planck2}
{Planck Collaboration} et~al., 2016, \mn@doi [\aap]
  {10.1051/0004-6361/201525830}, \href
  {http://adsabs.harvard.edu/abs/2016A%26A...594A..13P} {594, A13}

\bibitem[\protect\citeauthoryear{{Regan} \& {Downes}}{{Regan} \&
  {Downes}}{2018}]{rd18}
{Regan} J.~A.,  {Downes} T.~P.,  2018, \mn@doi [\mnras] {10.1093/mnras/sty134},
  \href {http://adsabs.harvard.edu/abs/2018MNRAS.475.4636R} {475, 4636}

\bibitem[\protect\citeauthoryear{{Regan} \& {Haehnelt}}{{Regan} \&
  {Haehnelt}}{2009}]{rh09b}
{Regan} J.~A.,  {Haehnelt} M.~G.,  2009, \mn@doi [\mnras]
  {10.1111/j.1365-2966.2009.14579.x}, \href
  {http://adsabs.harvard.edu/abs/2009MNRAS.396..343R} {396, 343}

\bibitem[\protect\citeauthoryear{{Rydberg}, {Whalen}, {Maturi}, {Collett},
  {Carrasco}, {Magg}  \& {Klessen}}{{Rydberg} et~al.}{2018}]{ryd18a}
{Rydberg} C.-E.,  {Whalen} D.~J.,  {Maturi} M.,  {Collett} T.,  {Carrasco} M.,
  {Magg} M.,   {Klessen} R.~S.,  2018, arXiv:1805.02662, \href
  {http://adsabs.harvard.edu/abs/2018arXiv180502662R} {}

\bibitem[\protect\citeauthoryear{{Sakurai}, {Hosokawa}, {Yoshida}  \&
  {Yorke}}{{Sakurai} et~al.}{2015}]{sak15}
{Sakurai} Y.,  {Hosokawa} T.,  {Yoshida} N.,   {Yorke} H.~W.,  2015, \mn@doi
  [\mnras] {10.1093/mnras/stv1346}, \href
  {http://adsabs.harvard.edu/abs/2015MNRAS.452..755S} {452, 755}

\bibitem[\protect\citeauthoryear{{Sakurai}, {Vorobyov}, {Hosokawa}, {Yoshida},
  {Omukai}  \& {Yorke}}{{Sakurai} et~al.}{2016}]{sak16b}
{Sakurai} Y.,  {Vorobyov} E.~I.,  {Hosokawa} T.,  {Yoshida} N.,  {Omukai} K.,
  {Yorke} H.~W.,  2016, \mn@doi [\mnras] {10.1093/mnras/stw637}, \href
  {http://adsabs.harvard.edu/abs/2016MNRAS.459.1137S} {459, 1137}

\bibitem[\protect\citeauthoryear{{Schauer}, {Regan}, {Glover}  \&
  {Klessen}}{{Schauer} et~al.}{2017}]{srg17}
{Schauer} A.~T.~P.,  {Regan} J.,  {Glover} S.~C.~O.,   {Klessen} R.~S.,  2017,
  \mn@doi [\mnras] {10.1093/mnras/stx1915}, \href
  {http://adsabs.harvard.edu/abs/2017MNRAS.471.4878S} {471, 4878}

\bibitem[\protect\citeauthoryear{{Shapiro} \& {Teukolsky}}{{Shapiro} \&
  {Teukolsky}}{1979}]{st79}
{Shapiro} S.~L.,  {Teukolsky} S.~A.,  1979, \mn@doi [\apjl] {10.1086/183134},
  \href {http://adsabs.harvard.edu/abs/1979ApJ...234L.177S} {234, L177}

\bibitem[\protect\citeauthoryear{{Smidt}, {Whalen}, {Johnson}, {Surace}  \&
  {Li}}{{Smidt} et~al.}{2018}]{smidt18}
{Smidt} J.,  {Whalen} D.~J.,  {Johnson} J.~L.,  {Surace} M.,   {Li} H.,  2018,
  \mn@doi [\apj] {10.3847/1538-4357/aad7b8}, \href
  {http://adsabs.harvard.edu/abs/2018ApJ...865..126S} {865, 126}

\bibitem[\protect\citeauthoryear{{Smith}, {Safranek-Shrader}, {Bromm}  \&
  {Milosavljevi{\'c}}}{{Smith} et~al.}{2015}]{aaron15}
{Smith} A.,  {Safranek-Shrader} C.,  {Bromm} V.,   {Milosavljevi{\'c}} M.,
  2015, \mn@doi [\mnras] {10.1093/mnras/stv565}, \href
  {http://adsabs.harvard.edu/abs/2015MNRAS.449.4336S} {449, 4336}

\bibitem[\protect\citeauthoryear{{Smith}, {Becerra}, {Bromm}  \&
  {Hernquist}}{{Smith} et~al.}{2017}]{aaron17}
{Smith} A.,  {Becerra} F.,  {Bromm} V.,   {Hernquist} L.,  2017, \mn@doi
  [\mnras] {10.1093/mnras/stx1993}, \href
  {http://adsabs.harvard.edu/abs/2017MNRAS.472..205S} {472, 205}

\bibitem[\protect\citeauthoryear{{Smith}, {Regan}, {Downes}, {Norman}, {O'Shea}
   \& {Wise}}{{Smith} et~al.}{2018}]{srd18}
{Smith} B.~D.,  {Regan} J.~A.,  {Downes} T.~P.,  {Norman} M.~L.,  {O'Shea}
  B.~W.,   {Wise} J.~H.,  2018, \mn@doi [\mnras] {10.1093/mnras/sty2103}, \href
  {http://adsabs.harvard.edu/abs/2018MNRAS.480.3762S} {480, 3762}

\bibitem[\protect\citeauthoryear{{Spergel} et~al.,}{{Spergel}
  et~al.}{2015}]{wfirst}
{Spergel} D.,  et~al., 2015, arXiv:1503.03757, \href
  {http://adsabs.harvard.edu/abs/2015arXiv150303757S} {}

\bibitem[\protect\citeauthoryear{{Spolyar}, {Freese}  \& {Gondolo}}{{Spolyar}
  et~al.}{2008}]{spoly08}
{Spolyar} D.,  {Freese} K.,   {Gondolo} P.,  2008, \mn@doi [\prl]
  {10.1103/PhysRevLett.100.051101}, \href
  {https://ui.adsabs.harvard.edu/abs/2008PhRvL.100e1101S} {100, 051101}

\bibitem[\protect\citeauthoryear{{Sun}, {Ruiz}  \& {Shapiro}}{{Sun}
  et~al.}{2018}]{sun18}
{Sun} L.,  {Ruiz} M.,   {Shapiro} S.~L.,  2018, \mn@doi [\prd]
  {10.1103/PhysRevD.98.103008}, \href
  {http://adsabs.harvard.edu/abs/2018PhRvD..98j3008S} {98, 103008}

\bibitem[\protect\citeauthoryear{{Surace} et~al.,}{{Surace}
  et~al.}{2018}]{sur18a}
{Surace} M.,  et~al., 2018, \mn@doi [\apjl] {10.3847/2041-8213/aaf80d}, \href
  {https://ui.adsabs.harvard.edu/abs/2018ApJ...869L..39S} {869, L39}

\bibitem[\protect\citeauthoryear{{Umeda}, {Hosokawa}, {Omukai}  \&
  {Yoshida}}{{Umeda} et~al.}{2016}]{um16}
{Umeda} H.,  {Hosokawa} T.,  {Omukai} K.,   {Yoshida} N.,  2016, \mn@doi
  [\apjl] {10.3847/2041-8205/830/2/L34}, \href
  {http://adsabs.harvard.edu/abs/2016ApJ...830L..34U} {830, L34}

\bibitem[\protect\citeauthoryear{{Valiante}, {Agarwal}, {Habouzit}  \&
  {Pezzulli}}{{Valiante} et~al.}{2017}]{rosa17}
{Valiante} R.,  {Agarwal} B.,  {Habouzit} M.,   {Pezzulli} E.,  2017, \mn@doi
  [Publications of the Astronomical Society of Australia]
  {10.1017/pasa.2017.25}, \href
  {http://adsabs.harvard.edu/abs/2017PASA...34...31V} {34, e031}

\bibitem[\protect\citeauthoryear{{Volonteri}, {Silk}  \& {Dubus}}{{Volonteri}
  et~al.}{2015}]{vsd15}
{Volonteri} M.,  {Silk} J.,   {Dubus} G.,  2015, \mn@doi [\apj]
  {10.1088/0004-637X/804/2/148}, \href
  {http://adsabs.harvard.edu/abs/2015ApJ...804..148V} {804, 148}

\bibitem[\protect\citeauthoryear{{Wang} et~al.,}{{Wang} et~al.}{2017}]{flare}
{Wang} L.,  et~al., 2017, arXiv:1710.07005, \href
  {http://adsabs.harvard.edu/abs/2017arXiv171007005W} {}

\bibitem[\protect\citeauthoryear{{Whalen} \& {Fryer}}{{Whalen} \&
  {Fryer}}{2012}]{wf12}
{Whalen} D.~J.,  {Fryer} C.~L.,  2012, \mn@doi [\apjl]
  {10.1088/2041-8205/756/1/L19}, \href
  {http://adsabs.harvard.edu/abs/2012ApJ...756L..19W} {756, L19}

\bibitem[\protect\citeauthoryear{{Whalen}, {Abel}  \& {Norman}}{{Whalen}
  et~al.}{2004}]{wan04}
{Whalen} D.,  {Abel} T.,   {Norman} M.~L.,  2004, \mn@doi [\apj]
  {10.1086/421548}, \href {http://adsabs.harvard.edu/abs/2004ApJ...610...14W}
  {610, 14}

\bibitem[\protect\citeauthoryear{{Whalen}, {Smidt}, {Johnson}, {Holz},
  {Stiavelli}  \& {Fryer}}{{Whalen} et~al.}{2013a}]{wet13c}
{Whalen} D.~J.,  {Smidt} J.,  {Johnson} J.~L.,  {Holz} D.~E.,  {Stiavelli} M.,
   {Fryer} C.~L.,  2013a, arXiv:1312.6330, \href
  {http://adsabs.harvard.edu/abs/2013arXiv1312.6330W} {}

\bibitem[\protect\citeauthoryear{{Whalen}, {Fryer}, {Holz}, {Heger}, {Woosley},
  {Stiavelli}, {Even}  \& {Frey}}{{Whalen} et~al.}{2013b}]{wet12a}
{Whalen} D.~J.,  {Fryer} C.~L.,  {Holz} D.~E.,  {Heger} A.,  {Woosley} S.~E.,
  {Stiavelli} M.,  {Even} W.,   {Frey} L.~H.,  2013b, \mn@doi [\apjl]
  {10.1088/2041-8205/762/1/L6}, \href
  {http://adsabs.harvard.edu/abs/2013ApJ...762L...6W} {762, L6}

\bibitem[\protect\citeauthoryear{{Whalen}, {Johnson}, {Smidt}, {Heger}, {Even}
  \& {Fryer}}{{Whalen} et~al.}{2013c}]{wet13b}
{Whalen} D.~J.,  {Johnson} J.~L.,  {Smidt} J.,  {Heger} A.,  {Even} W.,
  {Fryer} C.~L.,  2013c, \mn@doi [\apj] {10.1088/0004-637X/777/2/99}, \href
  {http://adsabs.harvard.edu/abs/2013ApJ...777...99W} {777, 99}

\bibitem[\protect\citeauthoryear{{Whalen}, {Smidt}, {Even}, {Woosley}, {Heger},
  {Stiavelli}  \& {Fryer}}{{Whalen} et~al.}{2014}]{wet13d}
{Whalen} D.~J.,  {Smidt} J.,  {Even} W.,  {Woosley} S.~E.,  {Heger} A.,
  {Stiavelli} M.,   {Fryer} C.~L.,  2014, \mn@doi [\apj]
  {10.1088/0004-637X/781/2/106}, \href
  {http://adsabs.harvard.edu/abs/2014ApJ...781..106W} {781, 106}

\bibitem[\protect\citeauthoryear{{Whalen}, {Hartwig}  \& {Surace}}{{Whalen}
  et~al.}{2019}]{til19a}
{Whalen} D.~J.,  {Hartwig} T.,   {Surace} M.,  2019, \mnras, in prep

\bibitem[\protect\citeauthoryear{{Windhorst} et~al.,}{{Windhorst}
  et~al.}{2018}]{wind18}
{Windhorst} R.~A.,  et~al., 2018, \mn@doi [\apjs] {10.3847/1538-4365/aaa760},
  \href {http://adsabs.harvard.edu/abs/2018ApJS..234...41W} {234, 41}

\bibitem[\protect\citeauthoryear{{Wise}, {Turk}  \& {Abel}}{{Wise}
  et~al.}{2008}]{wta08}
{Wise} J.~H.,  {Turk} M.~J.,   {Abel} T.,  2008, \mn@doi [\apj]
  {10.1086/588209}, \href {http://adsabs.harvard.edu/abs/2008ApJ...682..745W}
  {682, 745}

\bibitem[\protect\citeauthoryear{{Wise}, {Regan}, {O'Shea}, {Norman}, {Downes}
  \& {Xu}}{{Wise} et~al.}{2019}]{wise19}
{Wise} J.~H.,  {Regan} J.~A.,  {O'Shea} B.~W.,  {Norman} M.~L.,  {Downes}
  T.~P.,   {Xu} H.,  2019, \mn@doi [\nat] {10.1038/s41586-019-0873-4}, \href
  {http://adsabs.harvard.edu/abs/2019Natur.566...85W} {566, 85}

\bibitem[\protect\citeauthoryear{{Woods}, {Heger}, {Whalen}, {Haemmerl{\'e}}
  \& {Klessen}}{{Woods} et~al.}{2017}]{tyr17}
{Woods} T.~E.,  {Heger} A.,  {Whalen} D.~J.,  {Haemmerl{\'e}} L.,   {Klessen}
  R.~S.,  2017, \mn@doi [\apjl] {10.3847/2041-8213/aa7412}, \href
  {http://adsabs.harvard.edu/abs/2017ApJ...842L...6W} {842, L6}

\bibitem[\protect\citeauthoryear{{Woods} et~al.,}{{Woods}
  et~al.}{2018}]{titans}
{Woods} T.~E.,  et~al., 2018, arXiv:1810.12310, \href
  {http://adsabs.harvard.edu/abs/2018arXiv181012310W} {}

\bibitem[\protect\citeauthoryear{{Yorke} \& {Bodenheimer}}{{Yorke} \&
  {Bodenheimer}}{2008}]{yb08}
{Yorke} H.~W.,  {Bodenheimer} P.,  2008, in {Beuther} H.,  {Linz} H.,
  {Henning} T.,  eds,  Astronomical Society of the Pacific Conference Series
  Vol. 387, Massive Star Formation: Observations Confront Theory. p.~189

\bibitem[\protect\citeauthoryear{{Zackrisson} et~al.,}{{Zackrisson}
  et~al.}{2010a}]{z10b}
{Zackrisson} E.,  et~al., 2010a, \mn@doi [\mnras]
  {10.1111/j.1745-3933.2010.00908.x}, \href
  {http://adsabs.harvard.edu/abs/2010MNRAS.407L..74Z} {407, L74}

\bibitem[\protect\citeauthoryear{{Zackrisson} et~al.,}{{Zackrisson}
  et~al.}{2010b}]{z10a}
{Zackrisson} E.,  et~al., 2010b, \mn@doi [\apj] {10.1088/0004-637X/717/1/257},
  \href {http://adsabs.harvard.edu/abs/2010ApJ...717..257Z} {717, 257}

\makeatother
\end{thebibliography}







\bsp	
\label{lastpage}
\end{document}